\documentclass{article}

\usepackage{PRIMEarxiv}

\usepackage{graphicx}
\usepackage{makecell}
\usepackage{tabularx}
\usepackage{enumitem}
\usepackage{algorithm}
\usepackage{algpseudocode}
\usepackage{mathtools}
\usepackage{amsmath}
\usepackage{amssymb}
\usepackage{longtable,booktabs}
\usepackage{caption}
\usepackage{subcaption}
\usepackage{multirow}
\usepackage{cleveref}

\floatname{algorithm}{Procedure}

\title{A Survey of Multi-Agent Deep Reinforcement Learning with Communication}

\author{
  Changxi Zhu \\
  Department of Information and Computing Sciences \\
  Utrecht University \\
  Utrecht\\
  \texttt{c.zhu@uu.nl} \\
   \And
  Mehdi Dastani \\
  Department of Information and Computing Sciences \\
  Utrecht University \\
  Utrecht\\
  \texttt{m.m.dastani@uu.nl} \\
  \AND
  Shihan Wang \\
  Department of Information and Computing Sciences \\
  Utrecht University \\
  Utrecht \\
  \texttt{s.wang2@uu.nl} \\
}

\begin{document}

\maketitle

\begin{abstract}
Communication is an effective mechanism for coordinating the behaviors of multiple agents, broadening their views of the environment, and to support their collaborations. In the field of multi-agent deep reinforcement learning (MADRL), agents can improve the overall learning performance and achieve their objectives by communication. Agents can communicate various types of messages, either to all agents or to specific agent groups, or conditioned on specific constraints. With the growing body of research work in MADRL with communication (Comm-MADRL), there is a lack of a systematic and structural approach to distinguish and classify existing Comm-MADRL approaches. In this paper, we survey recent works in the Comm-MADRL field and consider various aspects of communication that can play a role in designing and developing multi-agent reinforcement learning systems. With these aspects in mind, we propose 9 dimensions along which Comm-MADRL approaches can be analyzed, developed, and compared. By projecting existing works into the multi-dimensional space, we discover interesting trends. We also propose some novel directions for designing future Comm-MADRL systems through exploring possible combinations of the dimensions.
\end{abstract}

% keywords can be removed

\section{Introduction}\label{sec1}

Many real-world scenarios, such as autonomous driving \cite{Shai2016MARLAD}, sensor networks \cite{Vinyals2011Sensor}, robotics \cite{Kober2013Robotics} and game-playing \cite{Silver2017Go,Brown2019superhuman}, can be modeled as multi-agent systems. Such multi-agent systems can be designed and developed using multi-agent reinforcement learning (MARL) techniques to learn the behavior of individual agents, which can be cooperative, competitive, or a mixture of them. As agents are often distributed in the environment where they only have access to their local observations rather than the complete state of the environment, partial observability becomes an essential assumption in MARL \cite{Oliehoek2016POMDP,Lowe2017MADDPG,Foerster2018COMA}. Moreover, MARL suffers from the non-stationary issue \cite{Papoudakis2019NonS}, since each agent faces a dynamic environment that can be influenced by the changing and adapting policies of other agents. Communication has been viewed as a vital means to tackle the problems of partial observability and non-stationary in MARL. Agents can communicate individual information, e.g., observations, intentions, experiences, or derived features, to have a broader view of the environment, which in turn allows them to make well-informed decisions \cite{Papoudakis2019NonS,Mohamed2019LearningComm}.

Due to the recent success of deep learning \cite{Lecun2015Deep} and its application to reinforcement learning \cite{Mnih2015DQN}, multi-agent deep reinforcement learning (MADRL) has witnessed great achievements in recent years, where agents can process high-dimensional data and have generalization ability in large state and action spaces \cite{Lowe2017MADDPG,Foerster2018COMA}. We notice that a large number of research works focus on \emph{learning tasks with communication}, which aim at learning to solve domain-specific tasks, such as navigation, traffic, and video games, by communicating and sharing information. To the best of our knowledge, there is a lack of survey literature that can cover recent works on learning tasks with communication in multi-agent deep reinforcement learning (Comm-MADRL). Early surveys consider the role of communication in MARL but assume it to be predefined rather than a subject of learning \cite{Stone2000Survey,Panait2005CMARL,Busoniu2008comprehensive}. Most Comm-MADRL surveys cover only a small number of research works without proposing a fine-grained classification system to compare and analyze them.\footnote{We provide a detailed comparison of recent surveys on MADRL which involves communication in Section \ref{section:recentsurveys}.} In cooperative scenarios, Hernandez-Leal et al. \cite{HernandezLeal2019Survey} use \textit{learning communication} to denote the area of learning communication protocols to promote the cooperation of agents.\footnote{In our survey, we extend the concept of \textit{learning communication} to general multi-agent tasks and use the term \emph{learning tasks with communication} to emphasize that the primary goal of recent research, which is centered on solving specific domain tasks through the use of communication.} The only survey that we found classifying some early works in Comm-MADRL is from Gronauer and Diepold \cite{Gronauer2021Survey}, which is based on distinguishing whether messages are received by all agents, a set of agents, or a network of agents. However, other aspects of Comm-MADRL, such as the type of messages and training paradigms, which are essential for communication and can help characterize existing communication protocols, are ignored. As a result, the reviewed papers in recent surveys regarding learning tasks with communication are rather limited and the proposed categorizations are too narrow to distinguish existing works in Comm-MADRL. On the other hand, there is a closely related research area, \emph{emergent language/communication}, which also considers learning communication through various reinforcement learning techniques \cite{EmergentComm2020}. Different from Comm-MADRL, the primary goal of emergent language studies is to learn a symbolic language.\footnote{In the literature, \emph{emergent language} and \emph{emergent communication} are used interchangeably. In our survey, we use \emph{emergent language} for referring to both terms.} However, a subset of emergent language research works pursues an additional goal to leverage learnable symbolic language to enhance task-level performance. Notably, these research works have not been encompassed within existing Comm-MADRL surveys but included in our survey, referred to \textit{learning tasks with emergent language}. In summary, our survey overlaps in scope with surveys of emergent language (i.e., in learning tasks with emergent language), but our survey focuses on different primary goals (i.e., achieving domain-specific tasks rather than learning a symbolic language). We further clarify the differences between learning tasks with communication and emergent language in Section \ref{section:extensions}.

In our survey paper, we review the Comm-MADRL literature by focusing on how communication can be utilized to improve the performance of multi-agent deep reinforcement learning techniques. Specifically, we focus on learnable communication protocols, which are aligned with recent works that emphasize the development of dynamic and adaptive communication, including learning when, how, and what to communicate with deep reinforcement learning techniques. Through a comprehensive review of recent Comm-MADRL literature, we propose a systematic and structured classification methodology designed to differentiate and categorize various Comm-MADRL approaches. Such a methodology will also provide guidance for the design and advancement of new Comm-MADRL systems. Suppose we plan to develop a Comm-MADRL system for a domain task at hand. Starting with the questions of when, how, and what to communicate, the system can be characterized from various aspects. Agents need to learn when to communicate, with whom to communicate, what information to convey, how to integrate received information, and, lastly, what learning objectives can be achieved through communication. We propose 9 dimensions that correspond to unique aspects of Comm-MADRL systems: Controlled Goals, Communication Constraints, Communicatee Type, Communication Policy, Communicated Messages, Message Combination, Inner Integration, Learning Methods, and Training Schemes. These dimensions, which form the skeleton of a Comm-MADRL system, can be used to analyze and gain insights into designed Comm-MADRL approaches thoroughly. By mapping recent Comm-MADRL approaches into this multi-dimensional structure, we not only provide insight into the current state of the art in this field but also determine some important directions for designing future Comm-MADRL systems. 

The remaining sections of this paper are organized as follows. In Section 2 the preliminaries of multi-agent RL are discussed, together with existing extensions regarding communication and a detailed comparison of recent surveys. In Section 3, we present our proposed dimensions, explaining how we group the recent works in the categories of each dimension. In Section 4, we discuss the trends that we found in the literature, and, driven by the proposed dimensions, we propose possible research directions in this research area. We finalize the paper with some conclusions in Section 5.

\section{Background}

In this section, we first provide the necessary background on multi-agent reinforcement learning. Then, we show how multi-agent reinforcement learning can be extended to consider communication between agents. Finally, we present and compare recent surveys involving communication, from which we can directly see our motivations to fill the gaps among existing surveys.

\subsection{Multi-agent Reinforcement Learning}

Real-world applications often contain more than one agent that operate in the environment. Agents are generally assumed to be autonomous and required to learn their strategies for achieving their goals. A multi-agent environment can be formalized in several ways \cite{Oliehoek2016DecPOMDP}, depending on whether the environment is fully observable, how agents' goals are correlated, etc. Among them, the Partially Observable Stochastic Game (POSG) \cite{Hansen2004POSG, Yang2020Survey} is one of the most flexible formalizations. A POSG is defined by a tuple $\left\langle\mathcal{I}, \mathcal{S},\rho^0,\left\{\mathcal{A}_{i}\right\}, P, \left\{\mathcal{O}_{i}\right\}, O, \left\{R_{i}\right\}\right\rangle$, where $\mathcal{I}$ is a (finite) set of agents indexed as $\{1,...,n\}$, $\mathcal{S}$ is a set of environment states, $\rho^{0}$ is the initial state distribution over state space $\mathcal{S}$, $\mathcal{A}_{i}$ is a set of actions available to agent $i$, and $\mathcal{O}_{i}$ is a set of observations of agent $i$. We denote a joint action space as $\boldsymbol{\mathcal{A}} =\times_{i\in \mathcal{I}}\mathcal{A}_{i}$ and a joint observation space of agents as $\boldsymbol{\mathcal{O}} =\times_{i\in \mathcal{I}}\mathcal{O}_{i}$. Therefore, $P: \mathcal{S} \times \boldsymbol{\mathcal{A}} \rightarrow \Delta(\mathcal{S})$ denotes the transition probability from a state $s \in \mathcal{S}$ to a new state $s' \in \mathcal{S}$ given agents' joint action $\vec{a} = \langle a_1,...,a_n \rangle$, where $\vec{a} \in \boldsymbol{\mathcal{A}}$. With the environment transitioning to the new state $s'$, the probability of observing a joint observation $\vec{o}=\langle o_1,...,o_n \rangle$ (where $\vec{o} \in \boldsymbol{\mathcal{O}}$) given the joint action $\vec{a}$ is determined according to the observation probability function $O: \mathcal{S} \times \boldsymbol{\mathcal{A}} \rightarrow \Delta(\boldsymbol{\mathcal{O}})$. Each agent then receives an immediate reward according to their own reward functions $R_i: \mathcal{S} \times \boldsymbol{\mathcal{A}} \times \mathcal{S} \rightarrow \mathbb{R}$. Similar to the joint action and observation, we could denote $\vec{r}=\langle r_1, ...,r_n \rangle$ as a joint reward. If agents' reward functions happen to be the same, i.e., they have identical goals, then $r_1=r_2=...=r_n$ holds for every time step. In this setting, the POSG is reduced to a Dec-POMDP \cite{Oliehoek2016DecPOMDP}. If at every time step the state is uniquely determined from the current set of observations of agents, i.e., $s\equiv\vec{o}$, the Dec-POMDP is reduced to a Dec-MDP. If each agent knows what the true environment state is, the Dec-MDP is reduced to a Multi-agent MDP. If there is only one single agent in the set of agents, i.e., $\mathcal{I}=\{1\}$, then the Multi-agent MDP is reduced to an MDP and the Dec-POMDP is reduced to a POMDP. Due to the partial observability, MARL methods often use the observation-action history $\tau_{i,t}=\{o_{i,0}, a_{i,0}, o_{i, 1},...,o_{i,t}\}$ up to time step $t$ for each agent to approximate the environment state. Note that time step $t$ is often omitted for the sake of simplification.  

In the multi-agent reinforcement learning setting, agents can learn their policies in either a decentralized or a centralized fashion. In decentralized learning (e.g., decentralized Q-learning \cite{Tan1993IL,Matignon2012Survey}), an $n$-agent MARL problem is decomposed into $n$ decentralized single-agent problems where each agent learns its own policy by considering all other agents as a part of the environment \cite{Claus1998Dynamics,Tampuu2015Pong}. In such a decentralized setting, the learned policy of each agent is conditioned on its local observation and history. A major problem with decentralized learning is the so-called non-stationarity of the environment, i.e., the fact that each agent learns in an environment where other agents are simultaneously exploring and learning. Centralized learning enables the training of either a single joint policy for all agents or a centralized value function to facilitate the learning of $n$ decentralized policies. While centralized (joint) learning removes or mitigates issues of partial observability and non-stationarity, it faces the challenge of joint action (and observation) spaces that expand exponentially with the number of agents and their actions. For a deeper dive into various training schemes used in MARL, we recommend the comprehensive survey by \cite{Gronauer2021Survey}, which offers valuable insights into the training and execution of policies. Based on whether policies are derived from value functions or directly learned, multi-agent reinforcement learning methods can be categorized into value-based and policy-based methods. Both methods have been largely utilized in Comm-MADRL.

\paragraph{Value-based}
Value-based methods in the multi-agent case borrow considerable ideas from the single-agent case. As one of the most popular value-based algorithms, the decentralized Q-learning learns a local Q-function for each agent. In the cooperative setting where agents share a common reward, the update rule for agent $i$ is as follows:
\begin{equation}
\label{eq:qlearningMul}
Q_i(s,a_i) \leftarrow Q_i(s,a_i) + \alpha (\underbrace{r + \gamma \max_{a'_i} Q_i(s',a'_i)}_{\text{new estimate}} - \underbrace{\vphantom{ \left(\frac{a^{0.3}}{b}\right) } Q_i(s,a_i)}_{\text{current estimate}})
\end{equation}
where $r$ is the shared reward, and $a'_i$ is the action with the highest Q-value in the next state $s'$. In partially observable environments, the environment state is not fully observable and is usually replaced by the individual observation or history of each agent. The Q-values for each state-action pair are incrementally updated according to the TD error. This error, i.e., $r + \gamma \max_{a'_i} Q_i(s',a'_i) - Q_i(s,a_i)$, represents the difference between a new estimate (i.e., $r + \gamma \max_{a'_i} Q_i(s',a'_i)$) and the current estimate (i.e., $Q_i(s,a_i)$) based on the Bellman equation \cite{Sutton2018RLBook}. As the state and action space could be too large to be encountered frequently for accurate estimation, function approximation methods, like deep neural networks, have become popular for endowing value or policy models with generalization abilities across both discrete and continuous states and actions \cite{Mnih2015DQN}. For example, the Deep Q-network (DQN) \cite{Mnih2015DQN} minimizes the difference between the new estimate calculated from sampled rewards and the current estimate of a parameterized Q-function. In DQN-based methods, the Q-function in Equation \ref{eq:qlearningMul} is notated as $Q_i(s,a_i;\theta_i)$, which depends on learnable parameters $\theta_i$. On the other hand, centralized learning in value-based methods learns a joint Q-function $Q(s, \vec{a}; \theta)$ with parameters $\theta$. However, this approach can be challenging to scale with an increasing number of agents. Value decomposition methods \cite{Sunehag2018VDN,Rashid2018Qmix,Son2019Qtran, Wang2021DOP} are popular MARL methods that decompose a joint Q-function to enable efficient training. These methods are also widely employed in research works in Comm-MADRL \cite{Zhang2019VBC,Zhang2020TMC,Yuan2022MAIC}. In partially observable environments, linear value decomposition methods decompose history-based joint Q-functions as follows:
\begin{equation}
\label{eq:decomposition}
Q^{joint}(\vec{\tau},\vec{a})= \sum_i^{n} w_i Q^i(\tau_i,a_i)
\end{equation}
where the joint Q-function is based on the joint history of all agents and is decomposed into local Q-functions based on individual histories. The weight $w_i$ can either be a fixed value \cite{Sunehag2018VDN,Son2019Qtran} or a learnable parameter subject to certain constraints \cite{Wang2021DOP}. Advantage functions can also replace the Q-function in the above equation to reduce variance \cite{Wang2021Qplex}.    

\paragraph{Policy-based}
Policy-based methods directly search over the policy space instead of obtaining the policy through value functions implicitly. The policy gradient theorem \cite{Sutton2018RLBook} provides an analytical expression of the gradients for a stochastic policy with learnable parameters in single-agent cases. In the multi-agent case with centralized learning, the policy gradient theorem is expressed as follows:
\begin{equation}
\label{eq:pgt}
\nabla_\theta J(\theta) = \mathbb{E}_{\vec{a}\sim \pi(\cdot \mid s),s\sim \rho^{\pi}}[\nabla_\theta \log \pi(\vec{a}\mid s;\theta)Q^{\pi}(s,\vec{a})]
\end{equation}
where $J(\theta)$ represents the learning objective, and $\pi(\vec{a}\mid s;\theta)$ denotes a stochastic policy parameterized by $\theta$ (abbreviated as $\pi$). Additionally, $\rho^{\pi}$ signifies the state distribution under the policy $\pi$, and $\nabla_\theta J(\theta)$ represents the expected gradient with respect to all possible actions and states. Due to the computational intractability of the expected gradient, stochastic gradient ascent can be applied to update the parameters $\theta$ at every learning step $l$ as follows:
$$
\theta_{l+1}=\theta_{l}+\alpha \widehat{\nabla_\theta J(\theta)}
$$
where $\alpha$ is the learning rate, and $\widehat{\nabla_\theta J(\theta)}$ is an estimate of the expected gradient based on sampled actions and states. Moreover, the Q-function in Equation \ref{eq:pgt} can be replaced by average returns over episodes to form REINFORCE algorithms \cite{Sutton2018RLBook}, or by an estimated value function to form actor-critic algorithms \cite{Konda1999AC,Schulman2016GAE}. In actor-critic methods, the policy and value function are termed the actor and the critic, respectively. The critic will, therefore, guide the learning of the actor.

Actor-critic methods have undergone various adaptations for multi-agent environments \cite{Lowe2017MADDPG,Foerster2018COMA,Oroojlooyjadid2019Survey,Papoudakis2020comparative}. A typical extension is the multi-agent deep deterministic policy gradient (MADDPG) \cite{Lowe2017MADDPG}. In MADDPG, the critic is a centralized Q-function designed to capture global information and coordinate learning signals. Meanwhile, the actors are local policies, ensuring decentralized execution. MADDPG assumes deterministic actors with continuous actions, allowing for the backpropagation of gradients from the value function to the policies. The gradient of each parameterized actor $\mu_{\theta_i}(a_i\mid o_i)$ with learnable parameters $\theta_i$, abbreviated as $\mu_i$, is defined as follows:
$$
\nabla_{\theta_{i}} J\left(\theta_i\right)=\mathbb{E}_{\vec{o}, \vec{a} \sim \mathcal{D}}\left[\nabla_{\theta_{i}} \mu_{i}\left(a_{i} \mid o_{i}\right) \nabla_{a_{i}} Q_{i}^{\mu}\left(\vec{o}, a_{1}, \ldots, a_{N}\right)\mid_{a_{i}=\mu_{i}\left(o_{i}\right)}\right]
$$
where $\mathcal{D}$ is the experience buffer that contains joint observation-action tuples $\langle \vec{o}, \vec{a}, \vec{r}, \vec{o'} \rangle$. Each agent's Q-function, denoted as $Q_{i}^{\mu}(\vec{o}, a_{1}, \ldots, a_{N})$, takes joint observations and actions as inputs, while decentralized actors use local observations as inputs. Contrary to Equation \ref{eq:pgt}, gradients with respect to the current action of agent $i$ (specifically, $\mu_{i}(o_{i})$) are utilized to guide the update of the policy parameter $\theta_{i}$. Both MADDPG and its single-agent counterpart, DDPG, have seen widespread application in Comm-MADRL \cite{Jiang2018ATOC,Malysheva2018MAGNet,Kilinc2018Noisy,Pesce2020MDMADDPG,Kim2019SchedNet}.

\subsection{Extensions with Communication}
\label{section:extensions}
\begin{figure}[t]
 \centering
 \includegraphics[width=0.5\linewidth]{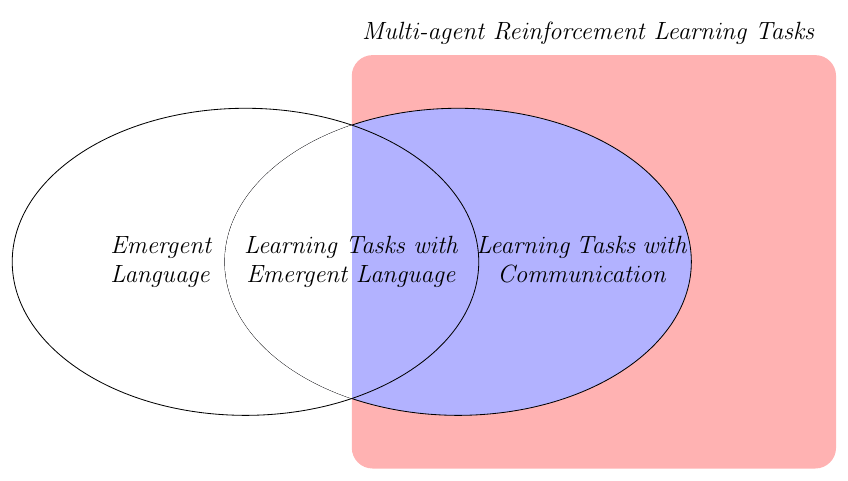}
 \caption{An illustration depicting the scope of this survey. The focus of our survey is represented by the blue part.}
 \label{fig:scope}
\end{figure}

In the MADRL literature where communication is used, we notice two closely related research areas, which we will refer to with the terms {\em emergent language} and {\em learning tasks with communication}. The {\em emergent language} research area \cite{EmergentComm2020,Negotiation2018Cao,Lowe2019Pitfalls,Quasi2021Bullard,Competition2021Noukhovitch} aims at learning a language grounded on symbols in communities of interacting/communicating agents. This line of research tries to understand the evolution of the language in agents equipped with neural networks. On the other hand, learning tasks with communication \cite{HernandezLeal2019Survey,Sukhbaatar2016CommNet,Singh2019IC3Net,Peng2017BicNet} focuses primarily on solving multi-agent reinforcement learning tasks with the aid of communication. Communication is often regarded as information exchange rather than learning a (human-like) language. Despite the distinction, when using MADRL techniques on specific domain tasks, languages might emerge, which can potentially enhance the learning system's explainability in accomplishing those tasks. We illustrate the research areas, emergent language and learning tasks with communication, along with their intersection \emph{learning tasks with emergent language} in Figure \ref{fig:scope}. Notably, our survey focuses on learning tasks with communication in multi-agent deep reinforcement learning, including the intersection with emergent language.\footnote{Throughout the remainder of our survey, Comm-MADRL will be used to specifically refer to the areas of our focus.} Within this focus, multiple agents often operate in partially observable environments and learn to share information encoded through neural networks. Furthermore, communication protocols, determining when and with whom to communicate, often leverage deep learning models to find the optimal choices that minimize communication overhead and yield more targeted communication. A multitude of works have been proposed to handle these subproblems inherent in Comm-MADRL. Most research works model only one or a few aspects of Comm-MADRL while selecting a default approach for other aspects. Given that the common goal of Comm-MADRL approaches is to design an effective and efficient communication protocol to improve agents’ learning performance in the environment, the proposed Comm-MADRL approaches inevitably share similarities to some extent. Consequently, establishing a classification system for Comm-MADRL becomes crucial. Such a system would aid in categorizing critical elements like contributions, targeted problems, and learning objectives, from which we can compare and analyse existing Comm-MADRL approaches. 

In the emergent language literature, numerous works employ various forms of the Lewis game, often referred to as referential games and operate under a cheap-talk setting \cite{farrell1996cheap}, as highlighted in several surveys \cite{Mohamed2019LearningComm, EmergentComm2020}.\footnote{In the emergent language research area, research works that do not adopt the cheap-talk setting but communicate through observable (domain-level) actions, are not included in our survey. Our survey focuses on explicit message transfer between agents.} In these games, a goal, often represented as a target location, an image, or a semantic concept, is given to a sender agent but remains unrevealed from a receiver agent. The receiver agent must then either identify the correct goal based on the sender's signaling \cite{Semantics2021Seo,Metropolis2022Taniguchi,Scale2022Chaabouni,Competition2021Noukhovitch,Compositionality2020Chaa,Capacity2020Resnick,Anti2019ChaabouniKDB,Emergence2017Havrylov} or accomplish its single-agent task using the received signals (messages) \cite{World2020Alexander,kajic2020learning}. Research works in learning tasks with emergent language are grounded in a multi-agent environment where the joint actions of both sender and receiver agents impact environment transitions. Consequently, the learning tasks with emergent language literature considers multi-agent domain tasks \cite{Populations2018Mordatch,Social2019Jaques,Biases2019Eccles,Discrete2021Tucker,Autoencoders2021Lin}, building on foundational concepts from MARL such as Dec-POMDPs or POSGs.

We further distinguish \textit{explicit} versus \textit{non-explicit} communication \cite{Oliehoek2016DecPOMDP} in the literature of MADRL with communication. Explicit communication refers to communication through a set of messages separate from domain-level actions. Here, agents' action policies are influenced by both their observations and the messages they receive. Such messages, crucial for supporting agents' decision-making, are essential in both the training and execution phases. MADRL frameworks without explicit communication can still allow for communication through domain-level actions, such as the act of influencing the observations of one agent through the actions of another. Furthermore, without explicit communication, agents can transmit gradient signals, which facilitate centralized training (and decentralized execution) but are not utilized during execution phases. Specifically, in our survey, we focus on explicit and learnable communication. 

Dec-POMDPs and POSGs are often extended to accommodate explicit communication. The communication can be integrated into the action set, adding a collection of communication acts alongside domain-level actions. Alternatively, a Dec-POMDP or a POSG can be extended to explicitly include a set of messages \cite{Oliehoek2016DecPOMDP}. For instance, the POSG can be expanded with a (shared) message space $\mathcal{M}$, resulting in a POSG-Comm, defined as $\left\langle\mathcal{I}, \mathcal{S},\rho^0,\left\{\mathcal{A}_{i}\right\}, P, \left\{\mathcal{O}_{i}\right\}, O, \left\{R_{i}\right\}, \mathcal{M} \right\rangle$, where all components remain unchanged except for the added message space $\mathcal{M}$. A Dec-POMDP-Comm can be defined as similar to the POSG-Comm with shared rewards. In both POSG-Comm and Dec-POMDP-Comm, action policies take into account both environmental observations and inter-agent messages. Research works in Comm-MADRL that expand upon a POSG or a Dec-POMDP can be seen in references such as \cite{Populations2018Mordatch,Biases2019Eccles,Wang2020IMAC,Autoencoders2021Lin,Xue2021MisSpoken}.

\subsection{Communication in Recent Surveys}
\label{section:recentsurveys}
Communication has attracted much attention in the field of multi-agent reinforcement learning (MARL). Previous surveys mentioning communication in MARL primarily focus on providing an overview of MARL's development. These surveys view communication as a subfield in MARL, and no extensive and substantial progress is reported. In an early survey, Stone and Veloso \cite{Stone2000Survey} classify MARL based on whether agents communicate and whether agents are homogeneous or not.\footnote{Homogeneous agents have the same internal structure including goals, domain knowledge, and possible actions.} They view learnable communication as a future research opportunity. Busoniu et al. \cite{Busoniu2008comprehensive} consider communication as a means to negotiate action choices and select equilibrium in the research direction of explicit coordination, without further classifying communication. With the advancement of deep learning, MARL has gradually incorporated deep neural networks such that recent developments are dominated by multi-agent deep reinforcement learning (MADRL). In the MADRL context, Hernandez-Leal et al. \cite{HernandezLeal2019Survey}, Nguyen et al. \cite{Nguyen2018Survey}, and Papoudakis et al. \cite{Papoudakis2019NonS} briefly review early Comm-MADRL methods, which have now become baselines in many recent works. Specifically, Hernandez-Leal et al. \cite{HernandezLeal2019Survey} use \textit{learning communication} to denote a new branch in MADRL. Papoudakis et al. \cite{Papoudakis2019NonS} consider communication as an approach to handle the non-stationary problem in MADRL, as agents can exchange information to stabilize their training. Compared to the aforementioned surveys, OroojlooyJadid and Hajinezhad \cite{Oroojlooyjadid2019Survey} provide a more detailed review of Comm-MADRL, covering a significant number of existing works. They view communication as a way to solve cooperative MADRL problems but did not propose a categorization model for Comm-MADRL. Zhang et al. \cite{Zhang2019Survey} and Yang et al. \cite{Yang2020Survey} review communication from a theoretical perspective. Their primary focus is on communication within networked multi-agent systems. In these systems, agents share information through a time-varying network, aiming to reach consensus on learned value functions or policies. Despite this, no further classification of communication is made.

Two more recent surveys in MADRL, proposed by Gronauer and Diepold \cite{Gronauer2021Survey} and Wong et al. \cite{Wong2021Survey}, focus on classifying existing works on communication. Gronauer and Diepold classify early research works in Comm-MADRL into Broadcasting, Targeted, and Networked communication, based on whether messages are received from all agents, a subset of agents, or a network of agents. Wong et al., similar to the survey of Papoudakis et al. \cite{Papoudakis2019NonS}, view communication as a method to address the issues of non-stationarity and partial observability. In the survey of Wong et al., research works on communication are categorized into three groups from a high-level perspective: communication as the primary learning goal, communication as an instrument to learn a specific task, and peer-to-peer teaching. However, they do not delve into how agents utilize communication to enhance learning. These surveys focus on limited aspects of communication, making their categorizations too narrow to distinguish recent works effectively, given the fact that many existing works share similar assumptions and conditions. To the best of our knowledge, only one survey \cite{Mohamed2019CommSurvey} exclusively focuses on communication issues in MADRL. They review algorithms for communication and cooperation, including efforts to interpret languages developed through communication. Despite this, their survey mainly covers early models without proposing a categorization framework.

The literature has investigated communication from other perspectives. Shoham and Leyton-Brown \cite{MAS2009Shoham} investigate communication from a game-theoretic perspective. They introduce several theories of communication in multi-agent systems, with the particular concern that agents can be self-motivated to convey information, driven by underlying incentives (e.g., the knowledge of game structure), or communicate in a pragmatic way analogous to human communication. Deep neural networks and deep reinforcement learning techniques have greatly widened the scope of language development in multi-agent systems. Lazaridou and Baroni \cite{EmergentComm2020} provide an extensive survey focused on $\emph{emergent language}$, aiming to establish effective human-machine communication. As highlighted in section \ref{section:extensions}, the primary goal of emergent language research is to learn a human-like language from scratch. The goal of our survey is, however, to classify the literature on learning tasks with communication that aims at exploiting communication to accomplish multi-agent tasks.

% \cite{Referential2018Lazaridou,Negotiation2018Cao,Competition2021Noukhovitch,Compositionality2020Chaa,Quasi2021Bullard}

In summary, existing surveys in Comm-MADRL lack coverage of the latest developments. These surveys also do not elaborate on the fact that communication itself is a combinatorial problem. Importantly, communication models engage with MADRL algorithms across various processes, including learning and decision-making. To effectively distinguish between existing Comm-MADRL approaches, it is crucial to analyze and classify them from a wider range of perspectives. In the following section, we delve into the field of Comm-MADRL through multiple dimensions, each linked to a unique research question pertinent to system design. These dimensions allow us to provide a fine-grained classification, highlighting the differences between Comm-MADRL approaches even within similar domains.

\section{Learning Tasks with Communication in MADRL}

In our survey, we consider explicit communication where action policies of agents are conditioned on communication that is learnable and dynamic, rather than static and predefined. Therefore, both the content of the messages and the chances of communication occurrences are subject to learning. As agents engage in multi-agent tasks, they learn domain-specific action policies and their communication protocols concurrently. As a result, \textit{learning tasks with communication} becomes a joint learning challenge, where agents employ reinforcement learning to maximize environmental rewards and simultaneously utilize various machine learning techniques to develop efficient and effective communication protocols.

Learning tasks with communication in multi-agent deep reinforcement learning (Comm-MADRL) is a significant research problem, particularly as communication can lead to higher rewards. Numerous studies have emerged, developing effective and efficient Comm-MADRL systems, often sharing similarities. Our review begins with the seminal works such as DIAL\cite{Foerster2016Comm}, RIAL\cite{Foerster2016Comm}, and CommNet\cite{Sukhbaatar2016CommNet}, and then expands to include the most relevant research works presented at major AI conferences and journals like AAMAS, AAAI, NeurIPS, and ICML, totaling 41 models in Comm-MADRL. To better distinguish among these models, we propose classifying them based on several dimensions in Comm-MADRL system design. These dimensions aim to comprehensively cover the current literature, allowing us to project the research works into a space where their similarities and differences become clear. We start by focusing on three key components of Comm-MADRL systems: problem settings, communication processes, and training processes. Problem settings encompass both communication-specific settings (e.g., communication constraints) and non-communication-specified settings (e.g., reward structures). Communication processes include common communication procedures, such as deciding whether to communicate and what messages to communicate. Training processes cover the learning of both agents and communication within MADRL. Based on the three key components, we identify and summarize 9 research questions that commonly arise in Comm-MADRL system design, corresponding to 9 dimensions as detailed in Table \ref{tab:step}. These research questions and dimensions are designed to capture various aspects of Comm-MADRL, covering the learning objectives of agents and communication, the processes by which messages are generated, transmitted, integrated, and learned within the MADRL framework. We outline a systematic procedure for providing a guideline to effectively navigate through these dimensions when developing Comm-MADRL systems. The procedure allows us to organize the dimensions, demonstrate their relevance in system design, and guide the creation of customized Comm-MADRL systems in a step-by-step manner.

\begin{table}[t]
  \caption{Proposed dimensions and associated research questions.}
  \label{tab:step}
  \begin{tabular}{p{0.2\linewidth}p{0.42\linewidth}p{0.2\linewidth}p{0.04\linewidth}}
  \hline
   \textit{Key Components} & \textit{Target Questions} & \textit{Dimensions} &  \textit{Index}
    \\\hline
    \multirow{3}{*}{Problem Settings}  & What kind of behaviors are desired to emerge with communication? & Controlled Goals & \textcircled{1} 
    \\\cline{2-4}
    & How to fulfill realistic requirements? & Communication Constraints & \textcircled{2} 
    \\\cline{2-4}
    & Which type of agents to communicate with? & Communicatee Type & \textcircled{3} 
    \\\hline
    \multirow{4}{*}{\begin{tabular}[c]{@{}c@{}}Communication \\ Processes\end{tabular}} & When and how to build communication links among agents? & Communication Policy & \textcircled{4} 
    \\\cline{2-4}
    & Which piece of information to share? & Communicated Messages & \textcircled{5} 
    \\\cline{2-4}
    & How to combine received messages? & Message Combination & \textcircled{6} 
    \\\cline{2-4}
    & How to integrate combined messages into learning models? & Inner Integration & \textcircled{7} 
    \\\hline
    \multirow{2}{*}{Training Processes} & How to train and improve communication? & Learning Methods & \textcircled{8}  
    \\\cline{2-4}
    & How to utilize collected experience from agents? & Training Schemes & \textcircled{9} 
    \\\hline
  \end{tabular}
\end{table}

As outlined in Procedure \ref{alg:prototype}, $N$ reinforcement learning agents employ communication throughout their learning and decision-making. Initially, the learning objective for the $N$ agents is set, defining rewards that induce cooperative, competitive, or mixed behaviors, as captured by dimension 1. We then consider potential communication-specified settings like limited resources, addressing the need for realistic scenarios as described in dimension 2. Dimension 3 identifies potential communicatees, determining the agents for messages to be received, which varies across domains. At each time step, agents decide when and with whom to communicate, as highlighted in dimension 4. The patterns of communication occurrences are structured like a graph, where links, either undirected or directed, aid information exchange. Subsequently, messages that encapsulate agents' understanding of the environment are generated and shared, relating to dimension 5. Given that agents often receive multiple messages, they must decide on how to combine these messages effectively. This process, crucial for integrating messages into their policies or value functions, is captured in dimensions 6 and 7. In cases of Comm-MADRL studies focusing on emergent language (i.e., learning tasks with emergent language), where messages are modeled as communicative acts emitted alongside domain-level actions, a specific rearrangement of the procedure is required. Here, messages are not observed by other agents until the next time step. Therefore, the processes outlined in dimensions 6 and 7 (lines 8 and 9) are moved to the front of those in dimension 4 (line 6). This rearrangement allows agents to combine and integrate messages from the previous time step before initiating new communication. As a result, agents make decisions and perform actions in the environment based not only on their environmental observations but also on information obtained from other agents (lines 10 and 11). During the training phase, experiences from both environmental interactions and inter-agent communication are utilized to train how agents will behave and communicate, i.e., agents' policies, value functions, and communication processes, as characterized in dimensions 8 and 9 (line 14).

In the following sections, we make an extensive survey on Comm-MADRL based on each dimension and classify the literature when we focus on a specific dimension. We finally provide a comprehensive table to frame recent works with the aid of the 9 dimensions.

\begin{algorithm}[h]
\caption{A guideline of Comm-MADRL systems}\label{alg:prototype}
\begin{algorithmic}[1]
\Require $N$ reinforcement learning agents
\State Set goals for reinforcement learning agents \Comment{Dimension \textcircled{1}}
\State Set possible communication constraints \Comment{Dimension \textcircled{2}}
\State Set the type of communicatees \Comment{Dimension \textcircled{3}}
\For{$episode= 1,2,...$}
\For{every environment step}
\State Decide with whom and whether to communicate \Comment{Dimension \textcircled{4}}
\State Decide which piece of information to share \Comment{Dimension \textcircled{5}}
\State Combine received information shared from others \Comment{Dimension \textcircled{6}}
\State Integrate messages into agents' internal models \Comment{Dimension \textcircled{7}}
\State Select actions based on communication
\State Perform in the environment (and store experiences)
\EndFor
\If{training is enablled}
\State Update agents' policies, value function, and communication processes \Comment{Dimensions \textcircled{8} \& \textcircled{9}}
\EndIf
\EndFor
\end{algorithmic}
\end{algorithm}
\captionof*{algorithm}{\textbf{Procedure 1}: A guideline of Comm-MADRL systems. The guideline positions dimensions where communication influences interaction with the environment and training phases.}

\subsection{Controlled Goal}

\begin{table}[t]
\caption{The category of controlled goals.}
\label{tab:goals}
\begin{tabular}{p{0.16\linewidth}p{0.20\linewidth}p{0.52\linewidth}}
    \toprule
    \small \textit{Types} & \small \textit{Configurations} & \small \textit{Methods}
    \\\midrule
    \small Cooperative & \small Global Rewards
    & 
DIAL \cite{Foerster2016Comm};
RIAL \cite{Foerster2016Comm};
CommNet \cite{Sukhbaatar2016CommNet};
GCL \cite{Populations2018Mordatch};
MAGNet-SA-GS-MG  \cite{Malysheva2018MAGNet};
MADDPG-M \cite{Kilinc2018Noisy};
SchedNet \cite{Kim2019SchedNet};
Agent-Entity Graph \cite{Agarwal2020AEG};
VBC \cite{Zhang2019VBC};
NDQ \cite{Wang2020NDQ};
IMAC \cite{Wang2020IMAC};
Gated-ACML \cite{Mao2020GatedACML};
Bias \cite{Biases2019Eccles};
LSC \cite{Sheng2020LSC};
Diff Discrete\cite{Freed2020UnknownNoise};
I2C \cite{Ding2020I2C};
TMC \cite{Zhang2020TMC};
GAXNet \cite{Yun2021GAXNet};
DCSS \cite{Discrete2021Tucker};
MAIC \cite{Yuan2022MAIC};
    \\\midrule
    & \small Local Rewards
    & 
BiCNet \cite{Peng2017BicNet};
DGN \cite{Jiang2020DGN};
IC3Net \cite{Singh2019IC3Net};
MD-MADDPG \cite{Pesce2020MDMADDPG};
DCC-MD \cite{Kim2019MessageDropout};
GA-Comm \cite{Liu2020G2ANet};
NeurComm \cite{Chu2020NeurComm};
IP \cite{Qu2020IP};
ETCNet \cite{Hu2020ETCNet};
Variable-length Coding \cite{Freed2020Length};
AE-Comm \cite{Autoencoders2021Lin};
    \\\midrule
    & \small Global or Local Rewards
    & 
MS-MARL-GCM \cite{Kong2017MSMARL};
ATOC \cite{Jiang2018ATOC};
TarMAC \cite{Das2019TarMAC};
IS \cite{Kim2021IS};
HAMMER \cite{Gupta2021HAMMER};
MAGIC \cite{Niu2021MAGIC};
FlowComm \cite{Du2021FlowComm};
FCMNet \cite{Wang2022FCMNet};
    \\\midrule
    \small Competitive  & \small Conflict Rewards
&
IC3Net \cite{Singh2019IC3Net};
R-MACRL \cite{Xue2021RMACRL};
    \\\midrule
    \small Mixed & \small Self-interested Rewards
&
IC \cite{Social2019Jaques};
DGN \cite{Jiang2020DGN};
TarMAC \cite{Das2019TarMAC};
IC3Net \cite{Singh2019IC3Net};
NDQ \cite{Wang2020NDQ};
LSC \cite{Sheng2020LSC};
MAGIC \cite{Niu2021MAGIC};
    \\\bottomrule
\end{tabular}
\end{table}

With a given reward configuration, reinforcement learning agents are guided to achieve their designated goals and interests. As agents communicate in order to obtain higher rewards, the goal of communication and the goal of achieving domain-specific tasks are inherently aligned. The emergent behaviors of agents can be summarized into three types: cooperative, competitive, and mixed \cite{Busoniu2006Survey,Matignon2012Survey}, each corresponding to different reward configurations and goals. Notably, some Comm-MADRL methods have been tested in more than one benchmark environment to show their flexibility and scalability, where the reward configurations may vary \cite{Singh2019IC3Net,Jiang2020DGN,Das2019TarMAC,Sheng2020LSC,Niu2021MAGIC}. Furthermore, a multi-agent environment may consist of both fixed opponents and teammates, which typically do not participate in communication. Therefore, we exclude fixed agents when identifying reward configurations. Consequently, we focus on (learnable) agents involved in communication and classify their behaviors that are desired to emerge, aligning them with associated reward configurations (summarized in Table \ref{tab:goals}).

\paragraph{\textbf{Cooperative}} In cooperative scenarios, agents have the incentive to communicate to achieve better team performance. Cooperative settings can be characterized by either a global reward that all agents share or a sum of local rewards that could be different among agents. Communication is usually used to promote cooperation as a team. Thus, in the literature, a team of agents can receive a global reward \cite{Foerster2016Comm,Sukhbaatar2016CommNet,Populations2018Mordatch,Kong2017MSMARL,Jiang2018ATOC,Das2019TarMAC,Malysheva2018MAGNet,Kilinc2018Noisy,Kim2019SchedNet,Agarwal2020AEG,Zhang2019VBC,Wang2020NDQ,Wang2020IMAC,Mao2020GatedACML,Biases2019Eccles,Sheng2020LSC,Freed2020UnknownNoise,Ding2020I2C,Kim2021IS,Zhang2020TMC,Gupta2021HAMMER,Niu2021MAGIC,Du2021FlowComm,Yun2021GAXNet,Discrete2021Tucker,Yuan2022MAIC,Wang2022FCMNet}, which does not account for the contribution of each agent. The agents can also receive local rewards, with designs to make the reward depend on teammates' collective performance \cite{Peng2017BicNet,Kong2017MSMARL,Jiang2020DGN,Singh2019IC3Net,Pesce2020MDMADDPG,Kim2019MessageDropout,Liu2020G2ANet,Freed2020Length,Gupta2021HAMMER}, to penalize collisions \cite{Jiang2018ATOC,Jiang2020DGN,Kim2019MessageDropout,Liu2020G2ANet,Kim2021IS,Hu2020ETCNet,Niu2021MAGIC,Du2021FlowComm}, or to share the reward with other agents for encouraging mutual cooperation \cite{Chu2020NeurComm,Qu2020IP,Autoencoders2021Lin}.

There are a variety of cooperative environments where communication has shown performance improvements, from small-scale games to complex video games. In early works, Foerster et al. \cite{Foerster2016Comm} developed two simple games, named Switch Riddle and MNIST Games, for their proposed models, DIAL and RIAL. Sukhbaatar et al. \cite{Sukhbaatar2016CommNet} used Traffic Junction for evaluating CommNet, which has become a popular testbed in recent works \cite{Kong2017MSMARL,Das2019TarMAC,Singh2019IC3Net,Liu2020G2ANet,Ding2020I2C,Kim2021IS,Niu2021MAGIC}. Among them, MAGIC \cite{Niu2021MAGIC} achieved higher performance on Traffic Junction with local rewards compared to two early works, CommNet \cite{Sukhbaatar2016CommNet}, IC3Net \cite{Singh2019IC3Net}, and one recent work, GA-Comm \cite{Liu2020G2ANet}. StarCraft \cite{Synnaeve2016TorchCraft,Vinyals2017StarCraftII,Samvelyan2019SMAC} is another benchmark environment in cooperative MARL with relatively flexible settings. BiCNet \cite{Peng2017BicNet} and MS-MARL-GCM \cite{Kong2017MSMARL} are evaluated on an early version of StarCraft \cite{Synnaeve2016TorchCraft}. Then, a new version of StarCraft, SMAC, has become popular in recent works \cite{Zhang2019VBC,Wang2020NDQ,Wang2020IMAC,Zhang2020TMC,Yuan2022MAIC,Wang2022FCMNet}. By controlling a team of agents, the cooperative goal in SMAC is to defeat enemies on easy, hard, and super hard maps. FCMNet \cite{Wang2022FCMNet} and MAIC are two recent works that surpass multiple communication methods and value decomposition methods (e.g., QMIX) on different maps. Google research football \cite{Kurach2020Football} is an even more challenging game with a physics-based 3D soccer simulator. Only MAGIC has reported performance on this platform with communication, and more investigations on this environment are needed. Compared to the above approaches in Comm-MADRL, ATOC \cite{Jiang2018ATOC} has been examined using a significantly larger number of learning agents in the predator-prey domain. Predator-prey is a grid world game with a long history in MARL. It has been developed with several versions \cite{Matignon2012IL,Brys2014PP,Lowe2017MADDPG}, while still viewed as a standard test environment due to its flexibility and customizability. ATOC reports performance on this platform with continuous state and action spaces. In the subfield \emph{learning tasks with emergent language}, cooperative scenarios are popularly used. They are mostly based on grid world or particle environments and have explicit role assignments, e.g., senders and receivers \cite{Populations2018Mordatch,Biases2019Eccles,Autoencoders2021Lin,Discrete2021Tucker}.

\paragraph{\textbf{Competitive}} In case agents need to compete with each other to occupy limited resources, they are assigned competitive learning objectives. In some competitive games, such as zero-sum games, one player wins and the others lose and therefore rational agents do not have the incentive to communicate. Nevertheless, in other competitive scenarios where agents compete for long-term goals, communication can allow for low-level cooperation among agents before the (long-term) goals are achieved. Based on our observations, only one work, IC3Net \cite{Singh2019IC3Net}, tests competitive settings and enables agents to compete for rewards.\footnote{IC3Net has been tested in several settings, including cooperative, competitive, and mixed scenarios, with different reward configurations.} IC3Net shows that competitive agents communicate only when it is profitable, e.g., before catching prey in the predator-prey domain. $\mathfrak{R}$-MACRL \cite{Xue2021RMACRL} considers communication from malicious agents to improve the worst-case performance. In $\mathfrak{R}$-MACRL, the whole environment is cooperative while agents learn to defend against malicious messages. Although the environment is cooperative, we classify this work under the competitive category as the learning goal between malicious agents and other agents is competitive.

\paragraph{\textbf{Mixed}} For a MAS where we care about self-interest agents, individual rewards can be designed and distributed to each agent \cite{Jiang2020DGN,Das2019TarMAC,Singh2019IC3Net,Wang2020NDQ,Sheng2020LSC,Niu2021MAGIC,Wang2022FCMNet}. Therefore, cooperative and competitive behaviors coexist during learning, which may show more complex communication patterns. Specifically, DGN \cite{Jiang2020DGN} considers a game where each agent gets positive rewards by eating food but gets higher rewards by attacking other agents. However, being attacked will get a high punishment. With communication, agents can learn to share resources collaboratively rather than attacking each other. IC3Net \cite{Singh2019IC3Net}, TarMAC \cite{Das2019TarMAC} and MAGIC \cite{Niu2021MAGIC} are evaluated on a mixed version of Predator-prey, and agents learn to communicate only when necessary. NDQ \cite{Wang2020NDQ} is examined in an independent search scenario, where two agents are rewarded according to their own goals, and shows that agents learn to not communicate in independent scenarios. IC \cite{Social2019Jaques} considers a scenario in which sender and receiver agents have different abilities to complete the goal. The sender agents have more vision but cannot clean obstacles, while receiver agents have limited vision but are able to clear obstacles. With communication, agents show collaborative behaviors to get higher rewards.

\subsection{Communication Constraints}
\label{section:CCCOM}

Practical concerns such as communication cost and noisy environment impair Comm-MADRL systems from embracing realistic applications more than simulations. This dimension, Communication Constraints, determines which type of communication concerns are handled in a Comm-MADRL system. We categorize recent works on this dimension into the following categories (summarized in Table \ref{tab:constraints}).

\paragraph{\textbf{Unconstrained Communication}}
In this category, communication processes, including communication channels, the content and transmission of messages, and the decisions of whether to communicate or not, are not explicitly restricted. In principle, agents can communicate as much as information they can without any decision to disallow communication in order to prevent communication overhead \cite{Sukhbaatar2016CommNet,Peng2017BicNet,Kong2017MSMARL,Jiang2020DGN,Das2019TarMAC,Malysheva2018MAGNet,Pesce2020MDMADDPG,Kim2021IS,Gupta2021HAMMER,Yun2021GAXNet,Wang2022FCMNet}. Specifically, several works consider blocking communication through predefined or learnable decisions of whether to communicate or not, while aiming to differentiate useful communicated information \cite{Jiang2018ATOC,Kilinc2018Noisy,Singh2019IC3Net,Kim2019MessageDropout,Agarwal2020AEG,Liu2020G2ANet,Sheng2020LSC,Chu2020NeurComm,Qu2020IP,Ding2020I2C,Niu2021MAGIC,Du2021FlowComm}. We also put those works under this category as they do not explicitly assume that communication is limited by cost.

\paragraph{\textbf{Constrained Communication}}
In this category, communication processes are explicitly constrained by cost or noise. Thus, agents need to utilize communication resources efficiently to promote learning. We further identify two practical concerns that have been considered in the literature.

\begin{table}[t]
\caption{The category of communication constraints.}
\label{tab:constraints}
\begin{tabular}{p{0.2\linewidth}p{0.13\linewidth}p{0.56\linewidth}}
    \toprule
    \small \textit{Types} & \small \textit{Subtypes} & \small \textit{Methods}
\\\midrule
\small Unconstrained Communication & &
CommNet \cite{Sukhbaatar2016CommNet};
BiCNet \cite{Peng2017BicNet};
MS-MARL-GCM \cite{Kong2017MSMARL};
ATOC \cite{Jiang2018ATOC};
DGN \cite{Jiang2020DGN};
TarMAC \cite{Das2019TarMAC};
MAGNet-SA-GS-MG  \cite{Malysheva2018MAGNet};
MADDPG-M \cite{Kilinc2018Noisy};
IC3Net \cite{Singh2019IC3Net};
MD-MADDPG \cite{Pesce2020MDMADDPG};
DCC-MD \cite{Kim2019MessageDropout};
Agent-Entity Graph \cite{Agarwal2020AEG};
GA-Comm \cite{Liu2020G2ANet};
LSC \cite{Sheng2020LSC};
NeurComm \cite{Chu2020NeurComm};
IP \cite{Qu2020IP};
I2C \cite{Ding2020I2C};
IS \cite{Kim2021IS};
HAMMER \cite{Gupta2021HAMMER};
MAGIC \cite{Niu2021MAGIC};
FlowComm \cite{Du2021FlowComm};
GAXNet \cite{Yun2021GAXNet};
FCMNet \cite{Wang2022FCMNet};
\\\midrule
\small Constrained Communication & Limited Bandwidth &
RIAL \cite{Foerster2016Comm};
DIAL \cite{Foerster2016Comm};
GCL \cite{Populations2018Mordatch};
IC \cite{Social2019Jaques};
SchedNet \cite{Kim2019SchedNet};
VBC \cite{Zhang2019VBC};
NDQ \cite{Wang2020NDQ};
IMAC \cite{Wang2020IMAC};
Gated-ACML \cite{Mao2020GatedACML};
Bias \cite{Biases2019Eccles};
ETCNet \cite{Hu2020ETCNet};
Variable-length Coding \cite{Freed2020Length};
TMC \cite{Zhang2020TMC};
AE-Comm \cite{Autoencoders2021Lin};
MAIC \cite{Yuan2022MAIC};
\\\midrule
\small & Corrupted Messages & 
DIAL \cite{Foerster2016Comm};
Diff Discrete\cite{Freed2020UnknownNoise};
DCSS \cite{Discrete2021Tucker};
$\mathfrak{R}$-MACRL \cite{Xue2021RMACRL};
\\\bottomrule
\end{tabular}
\end{table}

\begin{itemize}[leftmargin=*]
 \item \textbf{Limited Bandwidth}. In this category, communication bandwidth is limited by channel capacity. Thus, communication needs to be used more efficiently, both in the number of times that agents can communicate and the size of communicated information. Early works focus on transmitting succinct messages to avoid communication overhead. RIAL and DIAL \cite{Foerster2016Comm} are proposed to communicate very little information (i.e., a binary value or a real number) at every time step to reduce the bandwidth needed. MD-MADDPG \cite{Pesce2020MDMADDPG} considers a fixed-size memory, which is shared by all agents. Agents communicate through the shared memory instead of ad hoc channels. VBC \cite{Zhang2019VBC} and TMC \cite{Zhang2020TMC} reduce communication costs by using predefined thresholds to filter unnecessary communication, and both show lower communication overhead. NDQ \cite{Wang2020NDQ} cuts 80\% of messages by ordering the distributions of messages according to their means and drops accordingly to prevent meaningless messages. MAIC \cite{Yuan2022MAIC} also cuts messages by examining several message pruning rates. In MAIC, messages are encoded to consider their respective importance. Sent messages are ordered and then pruned with a given pruning rate. IMAC \cite{Wang2020IMAC} explicitly models bandwidth limitation as a constraint to optimization. An upper bound of the mutual information between messages and observations is derived according to bandwidth constraint, which turns out to minimize the entropy of messages. Then agents learn not only to maximize cumulative rewards but also to generate low-entropy messages. The number of agents to communicate can also be restricted to reduce the total amount of communication. SchedNet \cite{Kim2019SchedNet} considers a scenario of a shared channel together with limited bandwidth. Only a subset of agents are chosen to convey their messages according to their importance. Gated-ACML \cite{Mao2020GatedACML} learns a probabilistic gate unit to block messages transmitting between each agent and a centralized message coordinator, with the extra cost of learning optimal gates. Inspired by Gated-ACML and IMAC, ETCNet \cite{Hu2020ETCNet} puts constraints on the behaviors of deciding whether to send messages or not. A penalty term is added to the environment rewards, and an additional reinforcement learning algorithm is used to optimize the sending behaviors. Variable-length Coding \cite{Freed2020Length} also utilizes a penalty term while encouraging short messages. When learning tasks with emergent language, symbolic languages are acquired for communication through a limited number of tokens. Therefore, we classify those works under limited bandwidth \cite{Populations2018Mordatch,Social2019Jaques,Biases2019Eccles,Autoencoders2021Lin}.

\item \textbf{Corrupted Messages}. In this category, messages transmitted among agents can be corrupted due to environmental noise or malicious intentions. DIAL \cite{Foerster2016Comm} shows that during training, adding Gaussian noise to the communication channel can push the distribution of messages into two modes to convey different types of information. Diff Discrete \cite{Freed2020UnknownNoise} considers how to backpropagate gradients through a discrete communication channel (between 2 agents) with unknown noise. An encoder/channel/decoder system is modeled, where the encoder is used to discretize a real-valued signal into a discrete message to pass through the discrete communication channel, and the decoder is used to compute an approximation of the original signal. Later they show that the encoder/channel/decoder system is equivalent to an analog communication channel with additive noise. With the additional assumption that training is centralized, the gradient of the receiver with respect to real-value messages from the sender can be computed to allow backpropagation. DCSS \cite{Discrete2021Tucker} also considers a noisy setting. They prove that representing messages as one-hot vectors may not be optimal when the environment becomes noisy. Inspired by word embedding in the NLP field, they propose to generate a semantic representation of discrete tokens that are communicated among agents. The results show that such representation is robust in noisy environments and benefits human understanding of communication. Different from noisy environments, $\mathfrak{R}$-MACRL \cite{Xue2021RMACRL} assumes that an agent holds a malicious messaging policy, producing adversarial messages that can mislead other agents' action selections. Therefore, other agents need to prevent being exploited by learning a defense policy in order to filter the messages.   
\end{itemize}

\subsection{Communicatee Type}
\label{section:COMTYPE}

Communicatee Type determines which type of agents are assumed to receive messages in a Comm-MADRL system. We found that in the literature, communicatee type can be classified into the following categories based on whether agents in the environment communicate with each other directly or not.

\paragraph{\textbf{Agents in the MAS}} In this category, the set of communicatees consists of agents in the environment, and they directly communicate with each other. Nevertheless, due to partial observability, agents may not be able to communicate with every agent in the MAS, and thus we further distinguish the types of communicatees as follows:

\begin{itemize}[leftmargin=*]

\item \textbf{Nearby Agents}. In many Comm-MADRL systems, communication is only allowed between neighbors. Nearby agents can be defined as observable agents \cite{Yun2021GAXNet}, agents within a certain distance \cite{Jiang2020DGN,Agarwal2020AEG,Sheng2020LSC} or neighboring agents on a graph \cite{Chu2020NeurComm}. GAXNet \cite{Yun2021GAXNet} labels observable agents and enables communication between them. DGN \cite{Jiang2020DGN} limits communication within 3 closest neighbors while using a distance metric to find them. Agent-Entity Graph \cite{Agarwal2020AEG} also uses distance to measure whether agents are nearby or not. As long as two agents are close to each other, they will be allowed to communicate. LSC \cite{Sheng2020LSC} enables agents within a cluster radius to decide whether to become a leader agent. Then all non-leader agents in the same cluster will only communicate with the leader agent. NeurComm \cite{Chu2020NeurComm} and IP \cite{Qu2020IP} preset a graph structure among agents built upon networked multi-agent systems. In both NeurComm and IP, communicatees are restricted to neighbors on the graph. MAGNet-SA-GS-MG \cite{Malysheva2018MAGNet} uses a pre-trained graph to limit communication and restricts communication on neighboring agents. Neighboring agents can also emerge during learning instead of being predetermined, as proposed in GA-Comm \cite{Liu2020G2ANet}, MAGIC \cite{Niu2021MAGIC} and FlowComm \cite{Du2021FlowComm}, which explicitly learn a graph structure among agents. Specifically, in GA-Comm \cite{Liu2020G2ANet} and MAGIC \cite{Niu2021MAGIC}, a central unit (e.g., GNN) learns a graph inside and coordinates messages based on the (complete) graph simultaneously. In this case, agents do not communicate with each other directly; instead, they communicate through a virtual agent who does not affect the environment. Therefore, we categorize these two works into the proxy category.

\item \textbf{Other (Learning) Agents}. If nearby agents are not identified, the set of communicatees typically consists of other (learning) agents. Specifically, IC3Net \cite{Singh2019IC3Net} enables communication between learning agents and their opponents. Experiments indicate that these opponents eventually learn to not communicate to avoid being exploited. Some works assume explicit role assignments, i.e., senders and receivers. The role of the receiver can be taken by a disjoint set of agents separate from the senders \cite{Social2019Jaques,Biases2019Eccles,Discrete2021Tucker} or by all other agents in the environment \cite{Populations2018Mordatch,Autoencoders2021Lin}. In both cases, agents communicate with each other directly. 
\end{itemize}

\begin{table}[t]
\caption{The category of communicatee type.}
\label{tab:commitee}
\begin{tabular}{p{0.16\linewidth}p{0.16\linewidth}p{0.56\linewidth}}
    \toprule
    \small \textit{Types} & \small \textit{Subtypes} & \small \textit{Methods}
    \\\midrule
    \small Agents in the MAS & \small Nearby Agents &
DGN \cite{Jiang2020DGN};
MAGNet-SA-GS-MG  \cite{Malysheva2018MAGNet};
Agent-Entity Graph \cite{Agarwal2020AEG};
LSC \cite{Sheng2020LSC};
NeurComm \cite{Chu2020NeurComm};
IP \cite{Qu2020IP};
FlowComm \cite{Du2021FlowComm};
GAXNet \cite{Yun2021GAXNet};
\\\midrule
    & \small Other Agents &
DIAL \cite{Foerster2016Comm};
RIAL \cite{Foerster2016Comm};
CommNet \cite{Sukhbaatar2016CommNet};
GCL \cite{Populations2018Mordatch};
BiCNet \cite{Peng2017BicNet};
IC \cite{Social2019Jaques};
TarMAC \cite{Das2019TarMAC};
MADDPG-M \cite{Kilinc2018Noisy};
IC3Net \cite{Singh2019IC3Net};
SchedNet \cite{Kim2019SchedNet};
DCC-MD \cite{Kim2019MessageDropout};
VBC \cite{Zhang2019VBC};
NDQ \cite{Wang2020NDQ};
Bias \cite{Biases2019Eccles};
Diff Discrete\cite{Freed2020UnknownNoise};
I2C \cite{Ding2020I2C};
IS \cite{Kim2021IS};
ETCNet \cite{Hu2020ETCNet};
Variable-length Coding \cite{Freed2020Length};
TMC \cite{Zhang2020TMC};
AE-Comm \cite{Autoencoders2021Lin};
DCSS \cite{Discrete2021Tucker};
R-MACRL \cite{Xue2021RMACRL};
MAIC \cite{Yuan2022MAIC};
FCMNet \cite{Wang2022FCMNet};
\\\midrule
    \small Proxy & &
MS-MARL-GCM \cite{Kong2017MSMARL};
ATOC \cite{Jiang2018ATOC};
MD-MADDPG \cite{Pesce2020MDMADDPG};
IMAC \cite{Wang2020IMAC};
GA-Comm \cite{Liu2020G2ANet};
Gated-ACML \cite{Mao2020GatedACML};
HAMMER \cite{Gupta2021HAMMER};
MAGIC \cite{Niu2021MAGIC};
    \\\bottomrule
\end{tabular}
\end{table}

\paragraph{\textbf{Proxy}} A proxy is a virtual agent that plays an essential role (e.g., as a medium) in facilitating communication but does not directly affect the environment. Using a proxy as the communicatee means that agents will not directly communicate with each other, instead viewing the proxy as a medium, coordinating and transforming messages for specific purposes. MS-MARL-GCM \cite{Kong2017MSMARL} utilizes a master agent that collects local observations and hidden states from agents in the environment and sends a common message back to each of them. Similarly, HAMMER \cite{Gupta2021HAMMER} employs a central proxy that gathers local observations from agents and sends a private message to each agent. MD-MADDPG \cite{Pesce2020MDMADDPG} maintains a shared memory among agents, learning to selectively store and retrieve local observations from the memory. IMAC \cite{Wang2020IMAC} defines a scheduler that aggregates encoded information from all agents and sends individual messages to each agent. These works primarily focus on how to encode messages through the proxy without determining whether to send or receive messages. By contrast, ATOC \cite{Jiang2018ATOC}, Gated-ACML \cite{Mao2020CatedACML}, GA-Comm \cite{Liu2020G2ANet} and MAGIC \cite{Niu2021MAGIC} are all designed for agents to decide whether to communicate with a message coordinator. In ATOC and Gated-ACML, each agent's decisions are made locally based on individual observations, with messages aggregated from nearby agents and from the entire MAS, respectively. Both GA-Comm and MAGIC develop a global communication graph, coupled with a graph neural network (GNN) to aggregate messages by weights and send new messages back to each agent, informing action selection in the environment. \\

\begin{figure}[t]
 \centering
 \includegraphics[width=0.7\linewidth]{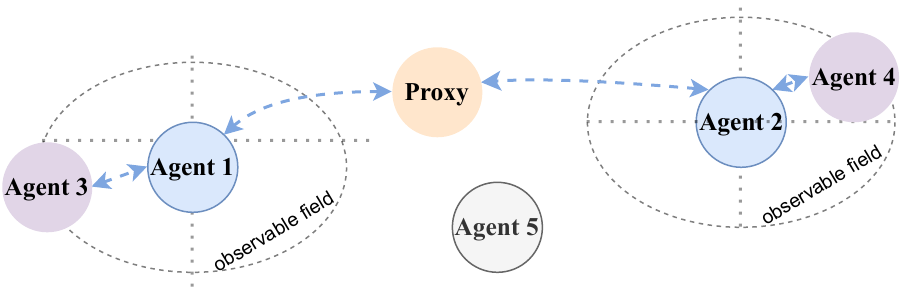}
 \caption{Three communicatee types in the same system.}
 \label{fig:commtype}
\end{figure}

Table \ref{tab:commitee} summarizes recent works on communication types in MAS. To illustrate these categories, we present an example of different communication methods used in a Comm-MADRL system in Figure \ref{fig:commtype}. The system consists of five agents and one proxy. Agent 3 is the nearby agent of Agent 1, while Agent 4 is the nearby agent of Agent 2. Agent 5 is out of the view range of Agents 1 and 2. If communication is limited to nearby agents, Agent 1 will communicate only with Agent 3, and Agent 2 will communicate only with Agent 4. However, if communication involves a proxy, all agents can send their messages to the proxy and receive coordinated messages.

\subsection{Communication Policy}
\label{section:COMPO}

Communication Policy determines when and with which agents (i.e., communicatees) to communicate in order to enable message transmission. A Communication Policy defines a set of communication actions, which can be modeled in different ways. For example, a communication action can be represented as a vector of binary values, where each value indicates whether communication with one of the other agents is allowed at a certain time step. These actions form communication links between pairs of agents, which can be represented as a communication graph among agents. In the literature, communication policies can be either predefined or learned, allowing communication with all other agents or only a subset of agents. Furthermore, communication policies can be centralized, controlling communication among all agents, or decentralized, enabling individual agents to control whether to communicate. Therefore, we first categorize the literature based on whether communication policies are predefined or learned. We find that in predefined communication policies, the literature often uses either full communication among agents, where the communication graph becomes complete, or a partial graph structure to incorporate constraints on communication policies. On the other hand, in learnable communication policies, we identify two distinct categories: individual control and global control. In individual control, communication policies are learned by each agent independently, whereas in global control, these policies are learned and implemented centrally, applying to all agents in Comm-MADRL systems. As a result, we have identified four subcategories within the dimension of communication policy: Full Communication, (Predefined) Partial Structure, Individual Control, and Global Control. These categorizations are summarized in Table \ref{tab:policy}.

\begin{figure}[t]
 \centering
%  \vskip 0pt
 \begin{subfigure}[t]{0.2\linewidth}
     \centering
     \includegraphics[width=0.9\linewidth]{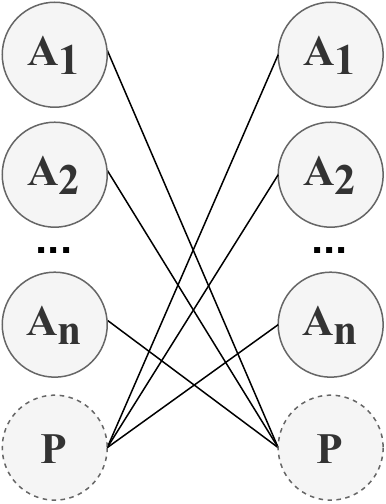}
     \label{fig:fullcomm}
    %  \caption{}
 \end{subfigure} \hspace*{\fill}
 \begin{subfigure}[t]{0.2\linewidth}
     \centering
     \includegraphics[width=0.9\linewidth]{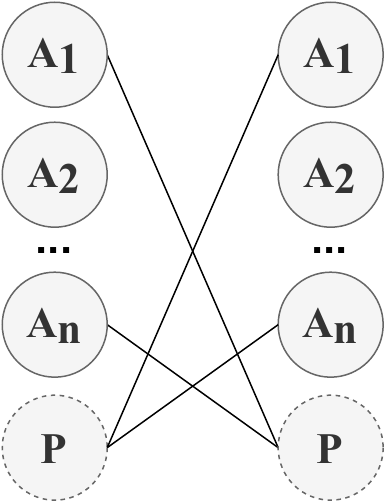}
     \label{fig:partialcomm}
    %  \caption{}
 \end{subfigure} \hspace*{\fill}
  \begin{subfigure}[t]{0.2\linewidth}
     \centering
     \includegraphics[width=0.9\linewidth]{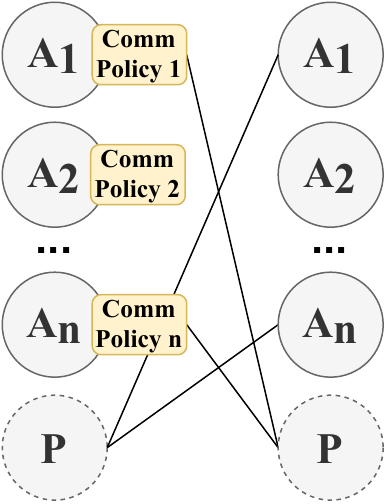}
     \label{fig:individual}
    %  \caption{}
  \end{subfigure} \hspace*{\fill}
  \begin{subfigure}[t]{0.2\linewidth}
     \centering
     \includegraphics[width=0.9\linewidth]{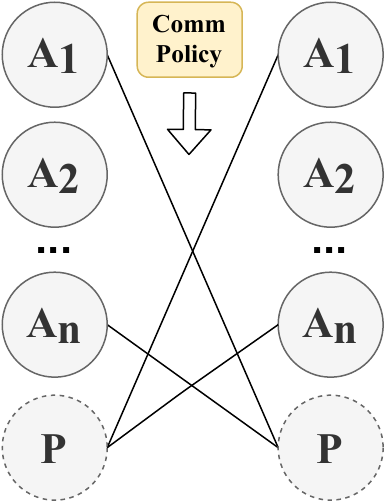}
     \label{fig:global} \hspace*{\fill}
    %  \caption{}
 \end{subfigure}
\caption{Four types of communication policy with agents (shown as A) in
the environment and a possible proxy (shown as P).}
\label{fig:fourgraphs}
\end{figure}

We present examples of how agents form communication links in the four categories of communication policy, as illustrated in Figure \ref{fig:fourgraphs}. Both Full Communication and Partial Structure rely on a predefined communication policy to determine communication actions. In contrast, Individual Control and Global Control involve the learning of a local communication policy and a global communication policy, respectively, to establish communication links between agents or a potential proxy. If a proxy is involved, it coordinates messages from agents choosing to communicate through this proxy. The categories and their associated research works are introduced as follows:

\begin{table}[t]
\caption{The category of communication policy}
\label{tab:policy}
\begin{tabular}{p{0.14\linewidth}p{0.20\linewidth}p{0.54\linewidth}}
    \toprule
    \small \textit{Types} & \small \textit{Subtypes} & \small \textit{Methods}
    \\\midrule
    \small Predefined & \small Full Communication & 
DIAL \cite{Foerster2016Comm};
RIAL \cite{Foerster2016Comm};
CommNet \cite{Sukhbaatar2016CommNet};
GCL \cite{Populations2018Mordatch};
BiCNet \cite{Peng2017BicNet};
MS-MARL-GCM \cite{Kong2017MSMARL};
TarMAC \cite{Das2019TarMAC};
MD-MADDPG \cite{Pesce2020MDMADDPG};
DCC-MD \cite{Kim2019MessageDropout};
IMAC \cite{Wang2020IMAC};
Diff Discrete\cite{Freed2020UnknownNoise};
IS \cite{Kim2021IS};
Variable-length Coding \cite{Freed2020Length};
HAMMER \cite{Gupta2021HAMMER};
AE-Comm \cite{Autoencoders2021Lin};
R-MACRL \cite{Xue2021RMACRL};
FCMNet \cite{Wang2022FCMNet};
\\\midrule
    & \small Partial Structure & 
IC \cite{Social2019Jaques};
DGN \cite{Jiang2020DGN};
MAGNet-SA-GS-MG  \cite{Malysheva2018MAGNet};
Agent-Entity Graph \cite{Agarwal2020AEG};
VBC \cite{Zhang2019VBC};
NDQ \cite{Wang2020NDQ};
Bias \cite{Biases2019Eccles};
NeurComm \cite{Chu2020NeurComm};
IP \cite{Qu2020IP};
TMC \cite{Zhang2020TMC};
GAXNet \cite{Yun2021GAXNet};
DCSS \cite{Discrete2021Tucker};
MAIC \cite{Yuan2022MAIC};
\\\midrule
    \small Learnable & \small Individual Control & 
ATOC \cite{Jiang2018ATOC};
MADDPG-M \cite{Kilinc2018Noisy};
IC3Net \cite{Singh2019IC3Net};
Gated-ACML \cite{Mao2020GatedACML};
LSC \cite{Sheng2020LSC};
I2C \cite{Ding2020I2C};
ETCNet \cite{Hu2020ETCNet};
\\\midrule
    & \small Global Control & 
SchedNet \cite{Kim2019SchedNet};
GA-Comm \cite{Liu2020G2ANet};
MAGIC \cite{Niu2021MAGIC};
FlowComm \cite{Du2021FlowComm};
    \\\bottomrule
\end{tabular}
\end{table}

\paragraph{\textbf{Full Communication}} In this category, every pair of agents is connected so that messages are transmitted in a broadcast manner. Full communication can be regarded as a fully connected graph, often used in early works on Comm-MADRL. DIAL \cite{Foerster2016Comm}, RIAL \cite{Foerster2016Comm}, CommNet \cite{Sukhbaatar2016CommNet}, and BiCNet \cite{Peng2017BicNet} learn a communication protocol which connect all agents together. Inspired by BiCNet, FCMNet \cite{Wang2022FCMNet} uses multiple RNNs to link all agents together with different sequences, allowing agents to benefit from communication flow from various directions. In contrast, Diff Discrete \cite{Freed2020UnknownNoise} and Variable-length Coding \cite{Freed2020Length} focus on two-agent cases but do not learn to block messages from each other. TarMAC \cite{Das2019TarMAC} and IS \cite{Kim2021IS} learn meaningful messages while using a broadcast way to share messages, thus still adhering to full communication. DCC-MD \cite{Kim2019MessageDropout} and $\mathfrak{R}$-MACRL \cite{Xue2021RMACRL} introduce a strategy to drop out received messages without specifying whether to send messages. Specifically, DCC-MD drops out messages with a fixed probability to reduce input dimensions, and $\mathfrak{R}$-MACRL learns to drop out adversary messages through a defense policy. In Comm-MADRL methods like IMAC \cite{Wang2020IMAC}, MS-MARL-GCM \cite{Kong2017MSMARL} and HAMMER \cite{Gupta2021HAMMER}, a central proxy that receives local observations or encoded messages is always connected with agents in the MAS. In addition, GCL \cite{Populations2018Mordatch} and AE-Comm \cite{Autoencoders2021Lin} learn a language grounded on discrete tokens among agents, where all agents have the capability to send and receive messages.

\paragraph{\textbf{(Predefined) Partial Structure}} In this category, the communication between agents is captured by a predetermined partial graph to reduce overall communication. Then, each agent communicates with a limited number of agents within the MAS, rather than with every agent. NeurComm \cite{Chu2020NeurComm} and IP \cite{Qu2020IP} operate in a networked multi-agent environment, randomly generating a communication network while maintaining a fixed average number of connections per agent during the learning process. DGN \cite{Jiang2020DGN}, MAGNet-SA-GS-MG \cite{Malysheva2018MAGNet}, and GAXNet \cite{Yun2021GAXNet} restrict communication to a certain proximity of agents. The Agent-Entity Graph \cite{Agarwal2020AEG} employs a pre-trained graph to capture agent relationships. Comm-MADRL approaches like VBC \cite{Zhang2019VBC}, NDQ \cite{Wang2020NDQ}, TMC \cite{Zhang2020TMC}, and MAIC \cite{Yuan2022MAIC} utilize handcrafted thresholds or pruning rates to limit communication opportunities. In IC \cite{Social2019Jaques}, Bias \cite{Biases2019Eccles}, and DCSS \cite{Discrete2021Tucker}, disjoint sets of agents are designated as either senders or receivers, facilitating unidirectional communication from senders to receivers only.

\paragraph{\textbf{Individual Control}} In this category, each agent actively and individually determines whether to communicate with other agents, implicitly forming a graph structure. A common method employed in Comm-MADRL studies within this category is a learnable gate mechanism, which aids agents in making the decision to communicate. For instance, IC3Net \cite{Singh2019IC3Net} and ATOC \cite{Jiang2018ATOC} use a gate mechanism that enables agents to decide whether to broadcast their messages, in a deterministic and probabilistic manner, respectively. ETCNet \cite{Hu2020ETCNet} also implements a gate unit but limits the overall probability of message-sending behaviors. If a proxy, such as a message coordinator, is present, Gated-ACML \cite{Mao2020GatedACML} introduces a learning mechanism for each agent to decide whether to communicate with the proxy, as opposed to direct communication with other agents. Diverging from the gate function approach, I2C \cite{Ding2020I2C} allows each agent to unilaterally decide on communication with other agents, based on evaluating the impact of those agents on its own policy. LSC \cite{Sheng2020LSC} allows each group of agents, defined by a specific radius, to compare their weights in order to elect a leader. This system then facilitates communication from each group to their respective leaders and from leader to leader. Notably, the leader agent in this model is not considered a proxy, as it still directly interacts with the environment.

\paragraph{\textbf{Global Control}} In this category, a globally shared communication policy is learned, providing more complete control over the communication links between agents. SchedNet \cite{Kim2019SchedNet} employs a global scheduler that limits the number of agents allowed to broadcast their messages, thereby reducing overall communication. FlowComm \cite{Du2021FlowComm} learns a directed graph among agents, enabling unilateral or bilateral communication between them. Similarly, GA-Comm \cite{Liu2020G2ANet} and MAGIC \cite{Niu2021MAGIC} develop an undirected and a directed graph for communication, respectively. These Comm-MADRL systems incorporate an additional message coordinator to coordinate and transform messages sent by the agents.

\subsection{Communicated Messages}
\label{section:COMMES}

After establishing communication links among agents through a communication policy, agents should determine which specific information to communicate. This information can derive from historical experiences, intended actions, or future plans, enriching the messages with valuable insights. Consequently, the communicated information can expand the agents' understanding of the environment and enhance the coordination of their behaviors. In the dimension of communicated messages, an important consideration is whether the communication includes future information, such as intentions and plans. This kind of information, being inherently private, often requires an (estimated) model of the environment to effectively simulate and generate conjectured intentions and plans. Accordingly, we categorize recent studies in this dimension into two categories, as summarized in Table \ref{tab:messages}.

\paragraph{\textbf{Existing Knowledge}} In this category, agents share their knowledge of the environment (e.g., past observations), previous movements, or policies to assist other agents in selecting actions. As historical information accumulates, agents use a low-dimensional encoding of their knowledge as messages to reduce communication overhead. Notably, the RNN family (e.g., LSTM and GRU) is commonly used as an encoding function, capable of selectively retaining and forgetting historical observations \cite{Sukhbaatar2016CommNet,Peng2017BicNet,Kong2017MSMARL,Das2019TarMAC,Singh2019IC3Net,Pesce2020MDMADDPG,Liu2020G2ANet,Ding2020I2C,Niu2021MAGIC,Du2021FlowComm,Yuan2022MAIC,Wang2022FCMNet}, action-observation histories \cite{Foerster2016Comm,Peng2017BicNet}, or action-observation-message histories \cite{Social2019Jaques,Biases2019Eccles}. When a proxy is present, messages are generated and transformed from agents to the proxy, and then from the proxy to agents. Thus, local observations can either be encoded \cite{Pesce2020MDMADDPG,Wang2020IMAC,Liu2020G2ANet,Mao2020GatedACML,Niu2021MAGIC} or directly sent \cite{Kong2017MSMARL,Gupta2021HAMMER} to the proxy. The proxy, after gathering these local (encoded) observations, can generate a unified message for all agents \cite{Kong2017MSMARL}, or individualized messages for each agent \cite{Pesce2020MDMADDPG,Wang2020IMAC,Liu2020G2ANet,Mao2020GatedACML,Gupta2021HAMMER,Niu2021MAGIC}. Both methods provide a message containing global information, relieving agents from the task of combining multiple received messages. In Comm-MADRL systems without a proxy, messages are sent directly to each agent. Specifically, in MADDPG-M \cite{Kilinc2018Noisy}, agents communicate local observations without an encoding of them. On the other hand, DIAL and RIAL \cite{Foerster2016Comm} encode past observations, actions, and current observations as messages. BiCNet \cite{Peng2017BicNet} encodes both local observations of each agent and a global view of the environment. Other research works employ various methods such as simple feed-forward networks \cite{Kim2019SchedNet,Freed2020UnknownNoise,Hu2020ETCNet,Freed2020Length}, MLP \cite{Malysheva2018MAGNet,Zhang2019VBC,Zhang2020TMC,Xue2021RMACRL}, autoencoders \cite{Kim2019MessageDropout}, CNNs \cite{Jiang2020DGN}, RNNs \cite{Sukhbaatar2016CommNet,Das2019TarMAC,Singh2019IC3Net,Ding2020I2C,Du2021FlowComm,Yuan2022MAIC,Wang2022FCMNet}, or GNNs \cite{Agarwal2020AEG,Sheng2020LSC} to encode local observations as messages. Furthermore, agents can communicate more specific information, such as in GAXNet \cite{Yun2021GAXNet}, where agents coordinate their local attention weights, integrating hidden states from neighboring agents. Messages can also be modeled as random variables, as seen in NDQ \cite{Wang2020NDQ}, where messages are drawn from a multivariate Gaussian distribution to maximize expressiveness by maximizing mutual information between messages and receivers' action selection. In learning tasks with emergent language, agents often communicate goal-related information, such as the goal's location \cite{Populations2018Mordatch,Social2019Jaques,Biases2019Eccles,Autoencoders2021Lin,Discrete2021Tucker}.

\begin{table}[t]
\caption{The category of communicated messages.}
\label{tab:messages}
\begin{tabular}{p{0.22\linewidth}p{0.7\linewidth}}
    \toprule
    \small \textit{Types} & \small \textit{Methods}
    \\\midrule
    \small Existing Knowledge & 
DIAL \cite{Foerster2016Comm};
RIAL \cite{Foerster2016Comm};
CommNet \cite{Sukhbaatar2016CommNet};
GCL \cite{Populations2018Mordatch};
BiCNet \cite{Peng2017BicNet};
MS-MARL-GCM \cite{Kong2017MSMARL};
IC \cite{Social2019Jaques};
DGN \cite{Jiang2020DGN};
TarMAC \cite{Das2019TarMAC};
MAGNet-SA-GS-MG  \cite{Malysheva2018MAGNet};
MADDPG-M \cite{Kilinc2018Noisy};
IC3Net \cite{Singh2019IC3Net};
MD-MADDPG \cite{Pesce2020MDMADDPG};
SchedNet \cite{Kim2019SchedNet};
DCC-MD \cite{Kim2019MessageDropout};
Agent-Entity Graph \cite{Agarwal2020AEG};
VBC \cite{Zhang2019VBC};
NDQ \cite{Wang2020NDQ};
IMAC \cite{Wang2020IMAC};
GA-Comm \cite{Liu2020G2ANet};
Gated-ACML \cite{Mao2020GatedACML};
Bias \cite{Biases2019Eccles};
LSC \cite{Sheng2020LSC};
Diff Discrete\cite{Freed2020UnknownNoise};
I2C \cite{Ding2020I2C};
ETCNet \cite{Hu2020ETCNet};
Variable-length Coding \cite{Freed2020Length};
TMC \cite{Zhang2020TMC};
HAMMER \cite{Gupta2021HAMMER};
MAGIC \cite{Niu2021MAGIC};
FlowComm \cite{Du2021FlowComm};
AE-Comm \cite{Autoencoders2021Lin};
GAXNet \cite{Yun2021GAXNet};
DCSS \cite{Discrete2021Tucker};
R-MACRL \cite{Xue2021RMACRL};
MAIC \cite{Yuan2022MAIC};
FCMNet \cite{Wang2022FCMNet};
    \\\midrule
    \small Imagined Future Knowledge & 
ATOC \cite{Jiang2018ATOC};
NeurComm \cite{Chu2020NeurComm};
IP \cite{Qu2020IP};
IS \cite{Kim2021IS};
    \\\bottomrule
\end{tabular}
\end{table}

%  ($\mathcal{I}$)
\paragraph{\textbf{Imagined Future Knowledge}} In this context, Imagined Future Knowledge refers to aspects such as intended actions \cite{Jiang2018ATOC}, policy fingerprints (i.e., action probabilities in a given state) \cite{Qu2020IP,Chu2020NeurComm}, or future plans \cite{Kim2021IS}. Since intentions are related to the current environment state, recent works often combine intended actions with local observations to produce more relevant messages. The concept of future plans extends this idea further by utilizing an approximated model of the environment and the behavior models of other agents. This approach enables the generation of a sequence of possible future observations and actions \cite{Kim2021IS}. Such knowledge is shared among agents, allowing the receivers to consider the potential future outcomes of the senders' actions.

\subsection{Message Combination}

\begin{table}[t]
\caption{The category of message combination.}
\centering
\begin{tabular}{p{0.22\linewidth}p{0.7\linewidth}}
    \toprule
    \small \textit{Types} & \small \textit{Methods}  
\\\midrule
    \small Equally Valued & 
DIAL \cite{Foerster2016Comm};
RIAL \cite{Foerster2016Comm};
CommNet \cite{Sukhbaatar2016CommNet};
GCL \cite{Populations2018Mordatch};
IC \cite{Social2019Jaques};
MADDPG-M \cite{Kilinc2018Noisy};
IC3Net \cite{Singh2019IC3Net};
SchedNet \cite{Kim2019SchedNet};
VBC \cite{Zhang2019VBC};
NDQ \cite{Wang2020NDQ};
Bias \cite{Biases2019Eccles};
Diff Discrete\cite{Freed2020UnknownNoise};
IS \cite{Kim2021IS};
ETCNet \cite{Hu2020ETCNet};
Variable-length Coding \cite{Freed2020Length};
FlowComm \cite{Du2021FlowComm};
AE-Comm \cite{Autoencoders2021Lin};
DCSS \cite{Discrete2021Tucker};
\\\midrule
    \small Unequally Valued &
BiCNet \cite{Peng2017BicNet};
MS-MARL-GCM \cite{Kong2017MSMARL};
ATOC \cite{Jiang2018ATOC};
DGN \cite{Jiang2020DGN};
TarMAC \cite{Das2019TarMAC};
MAGNet-SA-GS-MG  \cite{Malysheva2018MAGNet};
MD-MADDPG \cite{Pesce2020MDMADDPG};
DCC-MD \cite{Kim2019MessageDropout};
Agent-Entity Graph \cite{Agarwal2020AEG};
IMAC \cite{Wang2020IMAC};
GA-Comm \cite{Liu2020G2ANet};
Gated-ACML \cite{Mao2020GatedACML};
LSC \cite{Sheng2020LSC};
NeurComm \cite{Chu2020NeurComm};
IP \cite{Qu2020IP};
I2C \cite{Ding2020I2C};
TMC \cite{Zhang2020TMC};
HAMMER \cite{Gupta2021HAMMER};
MAGIC \cite{Niu2021MAGIC};
GAXNet \cite{Yun2021GAXNet};
R-MACRL \cite{Xue2021RMACRL};
MAIC \cite{Yuan2022MAIC};
FCMNet \cite{Wang2022FCMNet};
\\\bottomrule
\end{tabular}
\label{tab:combination}
\end{table}

When agents receive more than one message, current works often aggregate all received messages to reduce the input for the action policy. Message Combination determines how to integrate multiple messages before they are processed by an agent's internal model. If a proxy is involved, each agent receives already coordinated and combined messages from the proxy, eliminating the need for further message combination. If no proxy is presented, each agent independently determines how to combine multiple messages. Since communicated messages encode the senders' understanding of the learning process or the environment, some messages can be more valuable than others. As shown in Table \ref{tab:combination}, recent works in the dimension of message combination are categorized based on how agents prioritize received messages.

\paragraph{\textbf{Equally Valued}} In this category, messages received by agents are treated without preference, meaning they are assigned equal weights or simply no weights at all. Without having preferences, agents can concatenate all messages, ensuring no loss of information, though it may significantly expand the input space for the action policy \cite{Foerster2016Comm,Populations2018Mordatch,Social2019Jaques,Kilinc2018Noisy,Kim2019SchedNet,Wang2020NDQ,Freed2020UnknownNoise,Kim2021IS,Hu2020ETCNet,Freed2020Length}. Recent research involving concatenated messages typically represent the sent messages either as single values \cite{Foerster2016Comm,Social2019Jaques,Freed2020Length,Hu2020ETCNet} or as short vectors \cite{Populations2018Mordatch,Kilinc2018Noisy,Kim2019SchedNet,Wang2020NDQ,Freed2020UnknownNoise,Kim2021IS}. Alternatively, messages can be combined by averaging \cite{Sukhbaatar2016CommNet,Singh2019IC3Net,Zhang2019VBC} or summing \cite{Du2021FlowComm}, under the assumption that messages from different agents have the same dimension. In some cases, particularly in two-agent scenarios, no explicit preferences are assigned to messages \cite{Biases2019Eccles,Autoencoders2021Lin,Discrete2021Tucker}.

\paragraph{\textbf{Unequally Valued}} In this category, messages are assigned distinct preferences, which potentially impose differences on sender agents. DCC-MD \cite{Kim2019MessageDropout} and TMC \cite{Zhang2020TMC} use handcrafted rules to prune received messages. In DCC-MD, each received message can be dropped out with a certain probability. TMC stores the received messages and checks whether they are expired or not within a preset time window. Only valid messages are integrated into an agent's model. Instead of using fixed rules, $\mathfrak{R}$-MACRL \cite{Xue2021RMACRL} learns a gate unit to decide whether to use a received message. An attention mechanism can also be learned to assign weights to received messages and then combine them, rather than filtering messages out, as seen in research works \cite{Das2019TarMAC,Malysheva2018MAGNet,Agarwal2020AEG,Yuan2022MAIC}. Moreover, a neural network can aggregate received messages into a single message or a low-dimensional vector, which implicitly imposes preferences on messages during the mapping. Feedforward neural networks \cite{Wang2020IMAC,Mao2020GatedACML,Gupta2021HAMMER}, CNNs \cite{Jiang2020DGN}, LSTMs (or RNNs) \cite{Peng2017BicNet,Kong2017MSMARL,Jiang2018ATOC,Pesce2020MDMADDPG,Chu2020NeurComm,Ding2020I2C,Yun2021GAXNet,Wang2022FCMNet}, and GNNs \cite{Liu2020G2ANet,Sheng2020LSC,Qu2020IP,Niu2021MAGIC} have been used as aggregators. Among them, GNNs utilize a learned graph structure of agents and assign different weights to neighboring agents.

%  \cite{Velickovic2017GAT}
\begin{table}[t]
\caption{The category of inner integration.}
\centering
\begin{tabular}{p{0.22\linewidth}p{0.7\linewidth}}
    \toprule
    \small \textit{Types} & \small \textit{Methods}
\\\midrule
    \small Policy-level & 
CommNet \cite{Sukhbaatar2016CommNet};
GCL \cite{Populations2018Mordatch};
MS-MARL-GCM \cite{Kong2017MSMARL};
ATOC \cite{Jiang2018ATOC};
MAGNet-SA-GS-MG  \cite{Malysheva2018MAGNet};
IC3Net \cite{Singh2019IC3Net};
MD-MADDPG \cite{Pesce2020MDMADDPG};
SchedNet \cite{Kim2019SchedNet};
IMAC \cite{Wang2020IMAC};
GA-Comm \cite{Liu2020G2ANet};
Gated-ACML \cite{Mao2020GatedACML};
Diff Discrete\cite{Freed2020UnknownNoise};
IP \cite{Qu2020IP};
I2C \cite{Ding2020I2C};
IS \cite{Kim2021IS};
ETCNet \cite{Hu2020ETCNet};
Variable-length Coding \cite{Freed2020Length};
HAMMER \cite{Gupta2021HAMMER};
FlowComm \cite{Du2021FlowComm};
GAXNet \cite{Yun2021GAXNet};
R-MACRL \cite{Xue2021RMACRL};
\\\midrule
    \small Value-level & 
DIAL \cite{Foerster2016Comm};
RIAL \cite{Foerster2016Comm};
DGN \cite{Jiang2020DGN};
DCC-MD \cite{Kim2019MessageDropout};
VBC \cite{Zhang2019VBC};
NDQ \cite{Wang2020NDQ};
LSC \cite{Sheng2020LSC};
TMC \cite{Zhang2020TMC};
MAIC \cite{Yuan2022MAIC};
\\\midrule
    \small Policy- and Value-level &
BiCNet \cite{Peng2017BicNet};
IC \cite{Social2019Jaques};
TarMAC \cite{Das2019TarMAC};
MADDPG-M \cite{Kilinc2018Noisy};
Agent-Entity Graph \cite{Agarwal2020AEG};
Bias \cite{Biases2019Eccles};
NeurComm \cite{Chu2020NeurComm};
MAGIC \cite{Niu2021MAGIC};
AE-Comm \cite{Autoencoders2021Lin};
DCSS \cite{Discrete2021Tucker};
FCMNet \cite{Wang2022FCMNet};
\\\bottomrule
\end{tabular}
\label{tab:integration}
\end{table}

\subsection{Inner Integration}

Inner Integration determines how to integrate (combined) messages into an agent's learning model, such as a policy or a value function. In most existing literature, messages are viewed as additional observations. Agents take messages as extra input to a policy function, a value function, or both. Thus, in the dimension of inner integration, we classify recent works into categories based on the learning model that is used to integrate messages. These categories are summarized in Table \ref{tab:integration}.

%  ($\mathcal{P}_l$)
\paragraph{\textbf{Policy-level}} By exploiting information from other agents, each agent will no longer act independently. Policies can be learned through policy gradient methods like REINFORCE, as seen in studies \cite{Sukhbaatar2016CommNet,Populations2018Mordatch,Kong2017MSMARL,Singh2019IC3Net,Liu2020G2ANet}, which collect rewards during episodes and train the policy models at the end of episodes. Moreover, the Comm-MADRL approaches that utilize actor-critic methods \cite{Jiang2018ATOC,Malysheva2018MAGNet,Pesce2020MDMADDPG,Kim2019SchedNet,Wang2020IMAC,Mao2020GatedACML,Freed2020UnknownNoise,Qu2020IP,Ding2020I2C,Kim2021IS,Hu2020ETCNet,Freed2020Length,Gupta2021HAMMER,Du2021FlowComm,Yun2021GAXNet,Xue2021RMACRL} assume that a critic model (i.e., a Q-function) guides the learning of an actor model (i.e., a policy network).

% ($\mathcal{V}_l$)

\paragraph{\textbf{Value-level}} In this category, a value function incorporates messages as input, and a policy is derived by selecting the action with the highest Q-value. Most works in this category employ DQN-like methods to train their value functions \cite{Foerster2016Comm,Jiang2020DGN,Kim2019MessageDropout,Zhang2019VBC,Wang2020NDQ,Sheng2020LSC,Zhang2020TMC,Yuan2022MAIC}. Specifically, Comm-MADRL approaches like VBC \cite{Zhang2019VBC}, NDQ \cite{Wang2020NDQ}, TMC \cite{Zhang2020TMC}, and MAIC \cite{Yuan2022MAIC} are based on value decomposition methods in cooperative scenarios (with global rewards). These methods involve learning to decompose a joint Q-function.  

\paragraph{\textbf{Policy- and Value-level}} Integrating messages using both a policy function and a value function typically relies on actor-critic methods. In Comm-MADRL approaches within this category, received messages can be treated as extra inputs for both the actor and critic models \cite{Peng2017BicNet,Agarwal2020AEG,Discrete2021Tucker}. Alternatively, messages can be combined with local observations to generate new internal states, which are then shared with both the actor and critic models \cite{Social2019Jaques,Das2019TarMAC,Kilinc2018Noisy,Biases2019Eccles,Chu2020NeurComm,Niu2021MAGIC,Autoencoders2021Lin,Wang2022FCMNet}.

\subsection{Learning Methods}

Learning methods determine which type of machine learning techniques is used to learn a communication protocol. The learning of communication is at the center of modern Comm-MADRL and can benefit from the advancements in the machine learning field. If proper assumptions about communication are made, such as being able to calculate the derivatives with respect to the message generator function and the communication policy, then the training of communication can be integrated into the overall learning process of agents. This integration allows for the use of fully differentiable methods for backpropagation. Other machine learning techniques, including reinforcement learning, supervised learning, and regularizations, can also be utilized to incorporate our requirements and available ground truth into the learning of communication, each carrying respective assumptions. The assumptions used in the literature are summarized in Table \ref{tab:assumption}. For instance, supervised methods require defining true labels for communication (e.g., the correct information to share or the right agents to communicate with). In contrast, reinforced methods use rewards as learning signals. Regularized methods, which use neither true labels nor rewards, employ an additional learning objective by using regularizers, such as minimizing the entropy of messages to reduce stochasticity. Therefore, we classify recent works based on how they differ in the learning of communication (summarized in Table \ref{tab:learning}).

\begin{table}[t]
\caption{The assumptions behind different learning methods.}
\centering
\begin{tabular}{p{0.3\linewidth}p{0.6\linewidth}}
    \toprule
    \small \textit{Types} & \small \textit{Assumptions}
\\\midrule Fully differentiable & The messages or the communication actions are generated by a differentiable function and thus backpropagation is used everywhere.
\\\midrule Supervised learning & True labels (or the ground truth) are assumed to be given or defined to guide the learning of communication policy or messages.
\\\midrule Reinforcement learning & Environment rewards or self-defined rewards are used to update communication policy or messages incrementally.
\\\midrule Regularizers & Regularizations such as entropy inspired from information theory are added to agents' optimization objectives to regularize the learning of communication.
\\\bottomrule
\end{tabular}
\label{tab:assumption}
\end{table}

\begin{table}[t]
\caption{The category of learning methods.}
\label{tab:learning}
\begin{tabular}{p{0.2\linewidth}p{0.7\linewidth}}
    \toprule
    \small \textit{Types} & \small \textit{Methods} 
\\\midrule Differentiable &
GCL \cite{Populations2018Mordatch};
DIAL \cite{Foerster2016Comm};
CommNet \cite{Sukhbaatar2016CommNet};
BiCNet \cite{Peng2017BicNet};
MS-MARL-GCM \cite{Kong2017MSMARL};
DGN \cite{Jiang2020DGN};
TarMAC \cite{Das2019TarMAC};
MAGNet-SA-GS-MG  \cite{Malysheva2018MAGNet};
MD-MADDPG \cite{Pesce2020MDMADDPG};
DCC-MD \cite{Kim2019MessageDropout};
Agent-Entity Graph \cite{Agarwal2020AEG};
VBC \cite{Zhang2019VBC};
GA-Comm \cite{Liu2020G2ANet};
Diff Discrete\cite{Freed2020UnknownNoise};
NeurComm \cite{Chu2020NeurComm};
IP \cite{Qu2020IP};
IS \cite{Kim2021IS};
Variable-length Coding \cite{Freed2020Length};
TMC \cite{Zhang2020TMC};
MAGIC \cite{Niu2021MAGIC};
FlowComm \cite{Du2021FlowComm};
GAXNet \cite{Yun2021GAXNet};
DCSS \cite{Discrete2021Tucker};
FCMNet \cite{Wang2022FCMNet};

\\\midrule Supervised &
DCSS \cite{Discrete2021Tucker};
ATOC \cite{Jiang2018ATOC};
Gated-ACML \cite{Mao2020GatedACML};
I2C \cite{Ding2020I2C};
R-MACRL \cite{Xue2021RMACRL};

\\\midrule Reinforced & 
GCL \cite{Populations2018Mordatch};
RIAL \cite{Foerster2016Comm};
IC \cite{Social2019Jaques};
MADDPG-M \cite{Kilinc2018Noisy};
IC3Net \cite{Singh2019IC3Net};
SchedNet \cite{Kim2019SchedNet};
LSC \cite{Sheng2020LSC};
ETCNet \cite{Hu2020ETCNet};
HAMMER \cite{Gupta2021HAMMER};

\\\midrule Regularized & 
NDQ \cite{Wang2020NDQ};
IMAC \cite{Wang2020IMAC};
Bias \cite{Biases2019Eccles};
AE-Comm \cite{Autoencoders2021Lin};
MAIC \cite{Yuan2022MAIC};

\\\bottomrule
\end{tabular}
\end{table}

\paragraph{\textbf{Differentiable}} In this category, communication is learned and improved by backpropagating gradients from agent to agent. When the communication policy is predefined, such as in full communication \cite{Foerster2016Comm,Sukhbaatar2016CommNet,Populations2018Mordatch,Peng2017BicNet,Kong2017MSMARL,Das2019TarMAC,Pesce2020MDMADDPG,Kim2019MessageDropout,Freed2020UnknownNoise,Kim2021IS,Freed2020Length,Wang2022FCMNet} or by communicating with a subset of agents \cite{Jiang2020DGN,Malysheva2018MAGNet,Agarwal2020AEG,Zhang2019VBC,Chu2020NeurComm,Qu2020IP,Zhang2020TMC,Yun2021GAXNet,Discrete2021Tucker}, agents learn the content of messages through backpropagation. Several recent studies \cite{Populations2018Mordatch,Liu2020G2ANet,Niu2021MAGIC,Du2021FlowComm,Discrete2021Tucker} address the issue of non-differentiable communication actions by utilizing gradient estimators like Gumbel-softmax \cite{Jang2017Gum}, which replaces non-differentiable samples with a differentiable approximation during training, albeit requiring additional parameter tuning. Specifically, both GCL \cite{Populations2018Mordatch} and DCSS \cite{Discrete2021Tucker} employ a differential message function in their approaches. Additionally, GCL integrates auxiliary rewards, and DCSS utilizes labeled messages for training communication policies. Thus, they are categorized under the Differentiable category, each additionally aligning with the Reinforced and Supervised categories respectively. Freed et al. \cite{Freed2020UnknownNoise} propose an alternative method, Diff Discrete, to address the challenge of continuous messages versus discrete channels. This method models message transmitting as an encoder/channel/decoder system, where the receiver decodes the messages and reconstructs the original signals. These reconstructed signals enable the calculation of derivatives with respect to the sender, allowing gradients to be sent back to the sender.

\paragraph{\textbf{Supervised}} In this category, additional efforts need to be made to define the true label for when and what information to communicate. ATOC \cite{Jiang2018ATOC} and Gated-ACML \cite{Mao2020GatedACML} use the difference in Q-values between actions chosen with and without a message to define a label of communication actions. If the difference exceeds a threshold, the message is deemed valuable, indicating a high probability of sending it; otherwise, the probability is 0. This process sets up a classification task to decide whether to communicate. Similarly, I2C \cite{Ding2020I2C} trains a classifier to determine communication but relies on the causal effect between two agents, using a threshold to tag effective communication. $\mathfrak{R}$-MACRL \cite{Xue2021RMACRL} learns a classifier to identify malicious messages, using the status of a message (malicious or not) as a label. DCSS \cite{Discrete2021Tucker} learns message content by using a small dataset that maps observations to desired communication symbols. In DCSS, the gradient from the supervised loss is added to the policy loss, leading agents to use communication that aligns with the grounding data and enables high task performance.

\paragraph{\textbf{Reinforced}} In this category, reinforcement learning is utilized to train communication in addition to the learning of action policies. RIAL \cite{Foerster2016Comm} and HAMMER \cite{Gupta2021HAMMER} focus on learning the content of messages through reinforcement learning, without addressing the decision of whether to communicate. In GCL \cite{Populations2018Mordatch}, auxiliary rewards are used for predicting goals and consolidating symbols, facilitating the development of a compositional language for communication. IC \cite{Social2019Jaques} employs the difference in outcomes from using and not using communication on action policies as rewards. Maximizing the rewards can enhance the influence of communication on the receivers' action policies. Other studies \cite{Kilinc2018Noisy,Singh2019IC3Net,Kim2019SchedNet,Sheng2020LSC,Hu2020ETCNet} consider both the learning of communication content and the decision to communicate. Notably, MADDPG-M \cite{Kilinc2018Noisy} suggests using intrinsic rewards to train the communication policy instead of relying solely on environmental rewards. ETCNet \cite{Hu2020ETCNet} shapes environmental rewards by introducing a penalty term to discourage unnecessary communication.

\paragraph{\textbf{Regularized}} Regularized methods are used to reduce redundant information in communication \cite{Wang2020NDQ,Wang2020IMAC,Yuan2022MAIC}. NDQ \cite{Wang2020NDQ} calculates a lower bound of the mutual information between received messages and the receivers' action selection. This approach suggests that messages can be optimized to decrease the uncertainty in the action-value functions of the receivers. IMAC \cite{Wang2020IMAC} establishes an upper bound on the mutual information between messages and the senders' observations, and minimizing this upper bound helps agents send messages with lower uncertainty. MAIC \cite{Yuan2022MAIC} employs an estimated model of teammates and aims to maximize the mutual information between teammates' actions and hidden variables from this model. This model then guides the encoding of messages, resulting in tailored communications for different agents. Bias \cite{Biases2019Eccles} focuses on the long-term impact of messages on agents' decision-making to enhance signaling and listening effectiveness. AE-Comm \cite{Autoencoders2021Lin} adopts an autoencoder to learn a low-dimensional encoding of observations.

\subsection{Training Schemes}

This dimension focuses on how to utilize the collected experiences (such as observations, actions, rewards, and messages) of agents to train their action policies and communication architectures in a Comm-MADRL system. Agents can train their models in a fully decentralized manner using only their local experience. Alternatively, when global information is accessible, the experiences of all agents can be collected to centrally train a single (centralized) model that controls all agents. However, each approach has inherent challenges. Fully decentralized learning must cope with a non-stationary environment due to the changing and adapting behaviors of agents, while fully centralized learning faces the complexities of joint observation and policy spaces. As a balanced solution, Centralized Training and Decentralized Execution (CTDE) \cite{Kraemer2016CTDE,Foerster2016Comm} has emerged as a popular training schemes in MADRL. CTDE approaches allow agents to learn their local policies using guidance from central information. Therefore, in the dimension of training schemes, we categorize recent works based on how agents’ experiences are collected and utilized, as detailed in Table \ref{tab:scheme}.

\begin{table}[t]
\caption{The category of training schemes.}
\label{tab:scheme}
\begin{tabular}{p{0.18\linewidth}p{0.16\linewidth}p{0.54\linewidth}}
    \toprule
    \small \textit{Types} & \small \textit{Subtypes} & \small \textit{Methods}
\\\midrule
    \small Fully Decentralized Learning & &
IC \cite{Social2019Jaques};
MAGNet-SA-GS-MG  \cite{Malysheva2018MAGNet};
MADDPG-M \cite{Kilinc2018Noisy};
DCC-MD \cite{Kim2019MessageDropout};
Agent-Entity Graph \cite{Agarwal2020AEG};
Bias \cite{Biases2019Eccles};
NeurComm \cite{Chu2020NeurComm};
IP \cite{Qu2020IP};
AE-Comm \cite{Autoencoders2021Lin};
R-MACRL \cite{Xue2021RMACRL};
\\\midrule
    \small Centralized Training and Decentralized Execution & \small Individual Parameters &
MS-MARL-GCM \cite{Kong2017MSMARL};
SchedNet \cite{Kim2019SchedNet};
IMAC \cite{Wang2020IMAC};
Gated-ACML \cite{Mao2020GatedACML};
GAXNet \cite{Yun2021GAXNet};
DCSS \cite{Discrete2021Tucker};
\\\midrule
     & \small Parameter Sharing &
DIAL \cite{Foerster2016Comm};
RIAL \cite{Foerster2016Comm};
CommNet \cite{Sukhbaatar2016CommNet};
GCL \cite{Populations2018Mordatch};
BiCNet \cite{Peng2017BicNet};
ATOC \cite{Jiang2018ATOC};
DGN \cite{Jiang2020DGN};
TarMAC \cite{Das2019TarMAC};
IC3Net \cite{Singh2019IC3Net};
VBC \cite{Zhang2019VBC};
NDQ \cite{Wang2020NDQ};
GA-Comm \cite{Liu2020G2ANet};
LSC \cite{Sheng2020LSC};
Diff Discrete\cite{Freed2020UnknownNoise};
I2C \cite{Ding2020I2C};
ETCNet \cite{Hu2020ETCNet};
Variable-length Coding \cite{Freed2020Length};
TMC \cite{Zhang2020TMC};
HAMMER \cite{Gupta2021HAMMER};
MAGIC \cite{Niu2021MAGIC};
FlowComm \cite{Du2021FlowComm};
MAIC \cite{Yuan2022MAIC};
FCMNet \cite{Wang2022FCMNet};
\\\midrule
     & \small Concurrent &
MD-MADDPG \cite{Pesce2020MDMADDPG};
IS \cite{Kim2021IS};
\\\bottomrule
\end{tabular}
\end{table}

\begin{figure}
 \centering
 \begin{subfigure}[t]{0.48\linewidth}
     \centering
     \includegraphics[width=0.95\linewidth]{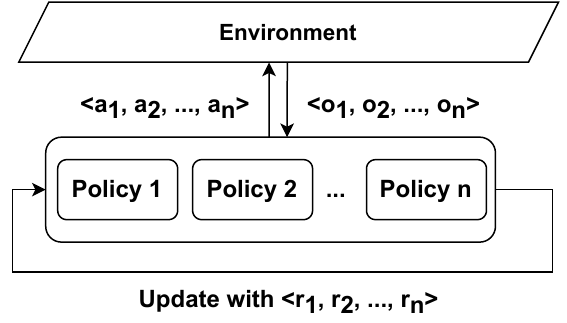}
     \caption{Fully Centralized Learning}
     \label{fig:CL}
 \end{subfigure}
 \begin{subfigure}[t]{0.48\linewidth}
     \centering
     \includegraphics[width=0.95\linewidth]{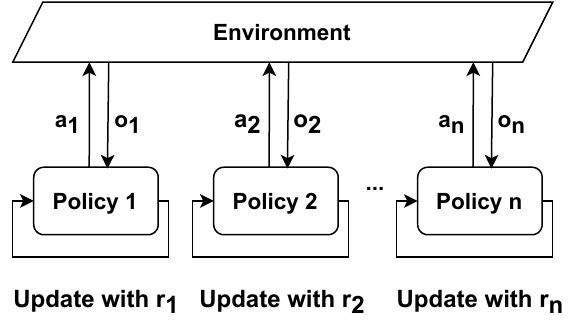}
     \caption{Fully Decentralized Learning}
     \label{fig:DL}
 \end{subfigure}
 
   \begin{subfigure}[t]{0.48\linewidth}
     \centering
     \includegraphics[width=0.95\linewidth]{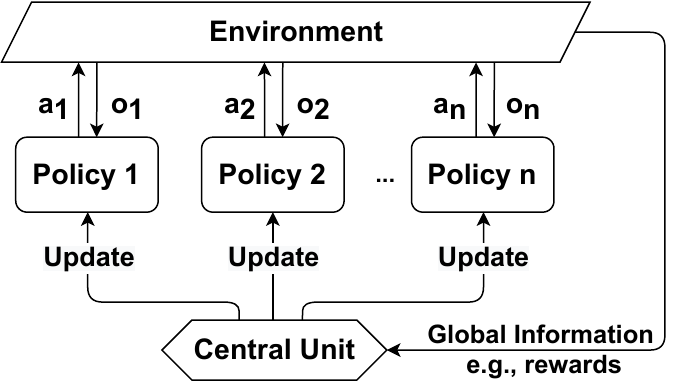}
     \caption{Individual Parameters}
     \label{fig:CTDEIP}
  \end{subfigure}
  \begin{subfigure}[t]{0.48\linewidth}
     \centering
     \includegraphics[width=0.95\linewidth]{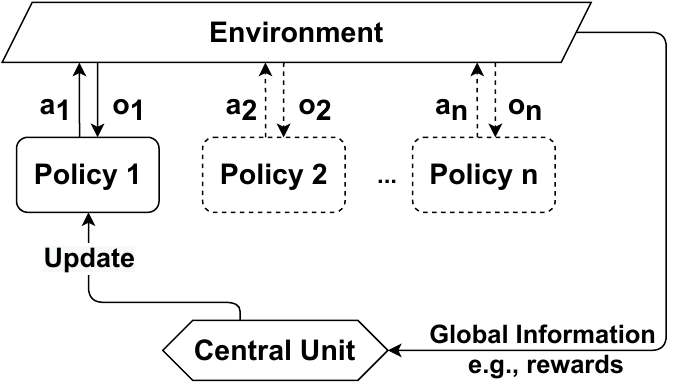}
     \caption{Parameter Sharing}
     \label{fig:CTDEPS}
 \end{subfigure}
 
  \begin{subfigure}[t]{\linewidth}
     \centering
     \includegraphics[width=0.45\linewidth]{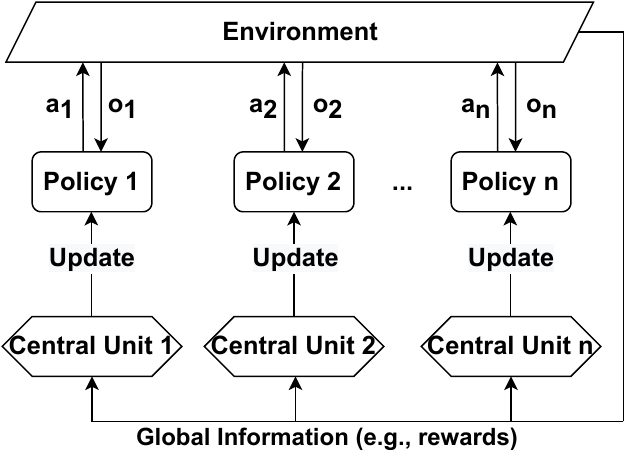}
     \caption{Concurrent}
     \label{fig:CCTDE}
 \end{subfigure}
\caption{Five types of Training Schemes.}
\label{fig:training}
\end{figure}

\paragraph{\textbf{Centralized Learning}} As shown in Figure \ref{fig:CL}, experiences are gathered into a central unit, then learning to control all agents. Based on our observations, recent works on Comm-MADRL usually do not assume a central controller.

\paragraph{\textbf{Fully Decentralized Learning}} As illustrated in Figure \ref{fig:DL}, in fully decentralized learning, experiences are collected individually by each agent, and they undergo independent training processes. Recent works in this category often employ actor-critic based methods for each agent \cite{Malysheva2018MAGNet,Kilinc2018Noisy,Kim2019MessageDropout,Agarwal2020AEG,Chu2020NeurComm,Qu2020IP,Xue2021RMACRL}. Specifically, decentralized learning has gained much attention in \emph{learning tasks with emergent language}, as it most closely resembles language learning in nature \cite{Social2019Jaques,Biases2019Eccles,Autoencoders2021Lin}.

\paragraph{\textbf{Centralized Training and Decentralized Execution}} In CTDE approaches, the experiences of all agents are collectively used for optimization. Gradients derived from the joint experiences of agents guide the learning of local policies. However, once training is complete, only the policies are needed and gradients can be discarded, facilitating decentralized execution. When agents are assumed to be homogeneous, meaning they have identical sensory inputs, actuators, and model structures, they can share parameters. Parameters sharing reduces the overall number of parameters, potentially enhancing learning efficiency compared to training in separate processes. Despite sharing parameters, agents can still exhibit distinct behaviors because they are likely to receive different observations at the same time step. Based on these considerations, recent works in this field can be further divided into the following subcategories.

\begin{itemize}[leftmargin=*]
\item \textbf{Independent Policies}. In this category, each local policy is trained with its own set of learning parameters. A central unit collects experiences from all agents to provide global information and guidance, such as gradients, as depicted in Figure \ref{fig:CTDEIP}. The training of the entire system can employ policy gradient algorithms (e.g., using REINFORCE) \cite{Kong2017MSMARL}, or actor-critic methods \cite{Kim2019SchedNet,Wang2020IMAC,Mao2020GatedACML,Yun2021GAXNet,Discrete2021Tucker}.

\item \textbf{Parameter Sharing}. In this category, all local policies (or local value functions) utilize a shared set of parameters, as illustrated in Figure \ref{fig:CTDEPS}. Commonly used algorithms in this scenario include DQN-like algorithms, actor-critic methods, and policy gradient algorithms with REINFORCE. When employing a DQN-like algorithm, a shared local Q-function, which processes each agent's individual experience, is learned collectively across agents \cite{Foerster2016Comm,Jiang2020DGN,Sheng2020LSC}. Additionally, DQN-based methods can be integrated with value decomposition models (e.g., QMIX \cite{Rashid2018Qmix}) in cooperative environments, which enable learning from factorized rewards (value functions) \cite{Zhang2019VBC,Wang2020NDQ,Zhang2020TMC,Yuan2022MAIC}. In the case of actor-critic methods, a shared actor (i.e., policy model) is trained using all individual experiences, supported by gradient guidance from a central critic \cite{Peng2017BicNet,Jiang2018ATOC,Das2019TarMAC,Freed2020UnknownNoise,Ding2020I2C,Hu2020ETCNet,Freed2020Length,Gupta2021HAMMER,Niu2021MAGIC,Du2021FlowComm,Wang2022FCMNet}. Policy gradient with REINFORCE can alternatively be used, requiring the collection of sampled rewards over episodes \cite{Sukhbaatar2016CommNet,Populations2018Mordatch,Singh2019IC3Net,Liu2020G2ANet}.

\item \textbf{Concurrent}. In scenarios where storing all experiences in a central unit is not feasible, agents can alternatively create backups of all experiences, with the assumption that they are able to observe other agents' actions and observations. The concurrent approaches differ inherently from fully decentralized learning. In CTDE with concurrent approaches, each agent maintains an individual set of policy parameters and receives the guidance from a local unit that collects global information (with additional assumptions on observability), as depicted in Figure \ref{fig:CCTDE}. Concurrent CTDE often employs actor-critic methods, where each agent has its own central critic to guide its local actor (policy) \cite{Pesce2020MDMADDPG,Kim2021IS}.
\end{itemize}

\subsection{Possible Relations of Dimensions}

We have introduced 9 dimensions for Comm-MADRL and identified a range of categories within each dimension. It is crucial to consider the potential interdependencies among these dimensions. We realize that the dimensions do not inherently depend on one another based on the criteria used for classifying the literature. However, specific implementations of Comm-MADRL systems may create dependencies between dimensions. For instance, limited bandwidth constraints (defined in the communication constraints dimension) can be realized by setting a limited number of times to communicate, rendering the full communication category (within the communication policy dimension) infeasible. This scenario illustrates how the dimensions of communication constraints (Section \ref{section:CCCOM}) and communication policy (Section \ref{section:COMPO}) become interdependent due to specific implementations. Another example about communicated messages shows that the classification criteria we used do not depend on each other. During implementation, a proxy (in the communicatee type dimension) or corrupted message constraints (in the communication constraints dimension) may change the value of message content. However, we categorize communicated messages as Existing Knowledge or Imagined Future Knowledge, based on whether future knowledge is simulated and utilized. This classification criterion is not inherently linked to a specific type of communicatee or communication constraint. Thus, the dimensions of communicatee type (Section \ref{section:COMTYPE}) and communication constraints (Section \ref{section:CCCOM}) are independent from the viewpoint of classification criteria. Consequently, the proposed categories and dimensions effectively encapsulate the literature from their unique perspectives.

\section{Findings, Discussions, and Research Directions}

In this section, we discuss the trend of the current literature and provide our observations and findings based on the proposed dimensions and categorizations. We also dive into the dimensions and suggest possible future research directions.

\subsection{Findings and Discussions}

To provide a more comprehensive overview of the literature, we have utilized the proposed 9 dimensions to categorize existing works, thereby creating an extensive table. For ease of reference, we introduce notations for these dimensions and their associated categories in Table \ref{tab:dimensions}. These notations are subsequently employed to categorize research works in Table \ref{tab:allworks}. In Table \ref{tab:allworks}, research works are sorted based on their publication or archival dates (e.g., on arXiv). Our proposed 9 dimensions offer different perspectives for analyzing and comparing recent works in the field of Comm-MADRL. Through these dimensions and categories, we have observed several intriguing findings.

\begin{table}[t]
\caption{The notations of all categories.}
\label{tab:dimensions}
\begin{tabular}{p{0.30\linewidth}p{0.62\linewidth}}
 \toprule
 \textit{Dimensions} & \textit{Notations} 
   \\\midrule
 Controlled Goals (CG) & $\mathcal{C}_{oo}$: Cooperative; $\mathcal{C}_{om}$: Competitive; $\mathcal{M}$: Mixed
 \\\midrule
  Communication Constraints (CC) &
  $\mathcal{U}$: Unconstrained Communication;$\mathcal{L}_b$: Limited Bandwidth; $\mathcal{C}_m$: Corrupted Messages;
  \\\midrule 
 Communicatee Type (CT) & $\mathcal{N}_a$: Nearby Agents; $\mathcal{A}$: Other (Learning) Agents; $\mathcal{P}$: Proxy
 \\\midrule
 Communication Policy (CP) & $\mathcal{F}_c$: Full Communication; $\mathcal{P}_s$: Predefined (Partial) Structure; $\mathcal{I}_c$: Individual Control; $\mathcal{G}_c$: Global Control
 \\\midrule
 Communicated Messages (CM) & $\mathcal{E}$: Existing Knowledge; $\mathcal{I}$: Imagined Future Knowledge
\\\midrule
 Message Combination (MC) & $\mathcal{V}_e$: Equally Valued; $\mathcal{V}_u$: Unequally Valued
  \\\midrule 
 Inner Integration (II) & $\mathcal{P}_l$: Policy-level; $\mathcal{V}_l$: Value-level; $\mathcal{PV}$: Policy-level \& Value-level
  \\\midrule
 Learning Methods (LM) & $\mathcal{D}$: Differentiable; $\mathcal{S}_p$: Supervised; $\mathcal{R}_e$: Reinforced; $\mathcal{R}_g$: Regularized
 \\\midrule
  Training Schemes (TS) & $\mathcal{CL}$: Centralized Learning; $\mathcal{DL}$: Decentralized Learning; $\mathcal{CTDE}_{ip}$: CTDE with Individual (Policy) Parameters; $\mathcal{CTDE}_{ps}$: CTDE with Parameter Sharing; $\mathcal{CTDE}_{c}$: Concurrent CTDE
 \\\bottomrule
\end{tabular}
\end{table}

\begin{itemize}[leftmargin=*]
\item In the dimension of Controlled Goals, recent research has focused on various cooperative settings, together with a few mixed scenarios. Communication in non-cooperative multi-agent tasks, however, has not been extensively explored. In such (non-cooperative) environments, the goals of different agents may conflict. In the emergent language literature, Noukhovitch et al. \cite{Competition2021Noukhovitch} have investigated how communication emerges between sender and receiver agents when they exhibit different levels of competitiveness, ranging from full cooperation to full competition. The results reveal that both sender and receiver agents can obtain higher rewards through communication when the level of competition is not high. However, their research primarily focuses on a simplified game without considering state transitions. The effectiveness of communication in MARL tasks with large state spaces, particularly in partial competitive settings where agents can still gain mutual benefits through low-level cooperation, remains an area for further exploration. Moreover, in non-cooperative settings, agents may be motivated to deceive or manipulate the communication channel to mislead others. The notion of \emph{trust} in multi-agent systems introduces the possibility of establishing a truthful communication protocol \cite{TrustSurvey2015Granatyr,Strategic2021Gunes}. Agents could assess the reliability of opponents and defend against malicious messages. Additionally, agents might evaluate interactions of opponents with other agents to determine their reputations, which could influence the priorities of communicating with other agents.

\item In the dimension of Communication Constraints, many existing works do not account for communication constraints, which may limit their applicability in realistic scenarios that have such limitations. For instance, transmitting messages in a large multi-agent system across long distances can result in delays, losses, or even be infeasible. Communication might be asynchronous, requiring several time steps for information exchange. These factors introduce new challenges to Comm-MADRL systems, such as validating previously sent messages and integrating messages from different time steps. Moreover, if communication resources are limited due to budget or capacity constraints, agents must decide how to allocate these resources effectively, especially when their goals vary. Conveying too much information might benefit others while sacrificing the agent's own learning opportunities. The concept of \emph{fairness}, which has been extensively studied in multi-agent systems, focuses on developing fair solutions for resource allocation. The ideas of maximizing the utility of worse-off agents and decreasing the difference in utilities between agents in the fairness study can be utilized to distribute communication resources equally. For instance, agents with lower utilities could be allotted more resources to facilitate their communication with others.

\item In the dimension of Communicatee Type, the concept of a proxy is utilized to facilitate message coordination. When global observability is available, a proxy often considers all agents within the environment. This proxy can be particularly effective and targeted by utilizing the independence among agents, coordinating messages among only a subset of agents as necessary.

\item In the dimension of Communication Policy, current works often assume a binary communication action regarding whether or not to communicate with other agents (or a specific agent). However, communication actions can be more fine-grained and descriptive. For instance, agents might opt to send only a portion of their messages due to uncertainty or lack of confidence. Additionally, a communication action could be defined more specifically, such as \emph{communicate with others if the budget exceeds a predetermined threshold}. Thus, a communication policy can encompass a variety of communication actions, tailored to align with human heuristics and specific system requirements.

\item In the dimension of Communicated Messages, various methods have been proposed to utilize the existing knowledge of agents for message generation. Some existing works consider incorporating agents' intentions or future plans. However, intentions or future plans may lead to catastrophic errors due to insufficient understanding of the underlying (transition) dynamics. Model-based Reinforcement Learning (RL) could assist agents in making more accurate predictions about future situations, thereby enabling the agents to communicate information with more certainty regarding upcoming changes. Additionally, current literature often assumes that messages are conveyed as single values or vectors. In contrast, modern devices allow for more complex formats, such as graphs and logical expressions. These formats can convey a substantial amount of knowledge or facts concisely, facilitating fast coordination. However, the challenge lies in effectively encoding and decoding complex information structures, which requires more sophisticated learning signals.

\item In the dimension of Message Combination, as messages often contain information related to each agent's individual experiences, goals, etc., many recent works consider the varying importance of these messages. These research works mostly rely on attention mechanisms to impose weights on received messages. Furthermore, agents can incorporate their prior knowledge or preferences about other agents’ capabilities into these weights, enhancing the relevance and effectiveness of message combination.

\item In the dimension of Inner Integration, many recent works have focused on integrating messages into the policy model. This trend is likely due to the growing interest in policy-based methods, particularly actor-critic algorithms, within the field of MADRL, where significant advancements have been achieved. Given that neural networks typically feature a hierarchical structure, there is potential for agents to effectively integrate messages into different layers. This approach would allow for considering varying levels of abstraction, potentially enhancing the decision-making process.

\item In the dimension of Learning Methods, the learning process for communication typically requires instantaneous feedback from agents who receive and act upon messages. This feedback could be in the form of gradient information or changes in the policies or rewards of the receiving agents. However, obtaining instantaneous feedback from other agents might not always be feasible in real-time decision-making systems. Despite this challenge, agents can still observe changes in the environment and their rewards to self-evaluate the effectiveness of their communication. This self-evaluation process enables agents to update and learn their communication protocols over time.

\item In the dimension of Training Schemes, parameter sharing combined with centralized training and decentralized execution is widely adopted in Comm-MADRL to reduce the number of learning parameters. However, accessing other agents' memories and parameters might raise privacy concerns. On the other hand, fully decentralized learning presents significant challenges and remains a key research area in MARL. In fully decentralized learning, agents have limited knowledge about the environment and must deal with non-stationarity, a problem that intensifies with an increasing number of agents. Nonetheless, Comm-MADRL can benefit from advancements in MARL, potentially leading to the development of novel training paradigms that better balance knowledge sharing, privacy, and learning efficiency.

\end{itemize}

Based on the proposed dimensions, we have identified a range of findings and potential issues in the field of Comm-MADRL. Among these issues, achieving fully decentralized learning and self-evaluated communication protocols remains a significant challenge. This difficulty arises because each agent has access only to their own data collected from the environment, adding complexity to message evaluation without the help of other agents. Decentralized action policies and self-evaluated communication protocols, however, could be advantageous in areas like Electronic Commerce \cite{Muller2014Survey}, Networks \cite{Manuel2020Networks}, and Blockchain \cite{Calvaresi2018Blockchain}, where synchronizing knowledge and information among users or agents can be computationally demanding. Another open question involves how to effectively communicate using more complex message formats and implement efficient training methods, potentially leading to more sophisticated communication architectures. Importantly, advancements in multi-agent systems and multi-agent reinforcement learning can significantly contribute to the progress of Comm-MADRL.

% \begin{small}
\begin{longtable}[t]{p{2.5cm}p{1cm}p{1cm}p{1cm}p{1cm}p{1cm}p{1cm}p{1cm}p{1cm}p{1cm}}
\caption{An overview of recent works in Comm-MADRL. $a+b$ denotes that the research work considers categories $a$ and $b$ simultaneously in the environment. $a/b$ denotes that the research work has been examined in multiple categories but in separate environments or settings.}
\label{tab:allworks}
  \\\toprule
    \textit{Methods} & CG & CC & CT & CP & CM & MC & II & LM & TS
    \\\midrule
\endfirsthead
\multicolumn{10}{r}{{Continued on next page}} \\
\endfoot
\endlastfoot
DIAL \cite{Foerster2016Comm}	&$\mathcal{C}_{oo}$	&$\mathcal{L}_b$+$\mathcal{C}_m$	&$\mathcal{A}$	&$\mathcal{F}_c$	&$\mathcal{E}$	&$\mathcal{V}_e$	&$\mathcal{V}_l$	&$\mathcal{D}$	&$\mathcal{CTDE}_{ps}$	\\\midrule
RIAL \cite{Foerster2016Comm}	&$\mathcal{C}_{oo}$	&$\mathcal{L}_b$	&$\mathcal{A}$	&$\mathcal{F}_c$	&$\mathcal{E}$	&$\mathcal{V}_e$	&$\mathcal{V}_l$	&$\mathcal{R}_e$	&$\mathcal{CTDE}_{ps}$	\\\midrule
CommNet \cite{Sukhbaatar2016CommNet}	&$\mathcal{C}_{oo}$	&$\mathcal{U}$	&$\mathcal{A}$	&$\mathcal{F}_c$	&$\mathcal{E}$	&$\mathcal{V}_e$	&$\mathcal{P}_l$	&$\mathcal{D}$	&$\mathcal{CTDE}_{ps}$	\\\midrule
GCL \cite{Populations2018Mordatch}	&$\mathcal{C}_{oo}$	&$\mathcal{L}_b$	&$\mathcal{A}$	&$\mathcal{F}_c$	&$\mathcal{E}$	&$\mathcal{V}_e$	&$\mathcal{P}_l$	&$\mathcal{D}$+$\mathcal{R}_e$	&$\mathcal{CTDE}_{ps}$	\\\midrule
BiCNet \cite{Peng2017BicNet}	&$\mathcal{C}_{oo}$	&$\mathcal{U}$	&$\mathcal{A}$	&$\mathcal{F}_c$	&$\mathcal{E}$	&$\mathcal{V}_u$	&$\mathcal{PV}$	&$\mathcal{D}$	&$\mathcal{CTDE}_{ps}$	\\\midrule
MS-MARL-GCM \cite{Kong2017MSMARL}	&$\mathcal{C}_{oo}$	&$\mathcal{U}$	&$\mathcal{P}$	&$\mathcal{F}_c$	&$\mathcal{E}$	&$\mathcal{V}_u$	&$\mathcal{P}_l$	&$\mathcal{D}$	&$\mathcal{CTDE}_{ip}$	\\\midrule
ATOC \cite{Jiang2018ATOC}	&$\mathcal{C}_{oo}$	&$\mathcal{U}$	&$\mathcal{P}$	&$\mathcal{I}_c$	&$\mathcal{I}$	&$\mathcal{V}_u$	&$\mathcal{P}_l$	&$\mathcal{S}_p$	&$\mathcal{CTDE}_{ps}$	\\\midrule
IC \cite{Social2019Jaques}	&$\mathcal{C}_{oo}$	&$\mathcal{L}_b$	&$\mathcal{A}$	&$\mathcal{F}_c$	&$\mathcal{E}$	&$\mathcal{V}_e$	&$\mathcal{PV}$	&$\mathcal{R}_e$	&$\mathcal{DL}$	\\\midrule
DGN \cite{Jiang2020DGN}	&$\mathcal{C}_{oo}$/$\mathcal{M}$	&$\mathcal{U}$	&$\mathcal{N}_{a}$	&$\mathcal{P}_s$	&$\mathcal{E}$	&$\mathcal{V}_u$	&$\mathcal{V}_l$	&$\mathcal{D}$	&$\mathcal{CTDE}_{ps}$	\\\midrule
TarMAC \cite{Das2019TarMAC}	&$\mathcal{C}_{oo}$/$\mathcal{M}$	&$\mathcal{U}$	&$\mathcal{A}$	&$\mathcal{F}_c$	&$\mathcal{E}$	&$\mathcal{V}_u$	&$\mathcal{PV}$	&$\mathcal{D}$	&$\mathcal{CTDE}_{ps}$	\\\midrule
MAGNet-SA-GS-MG  \cite{Malysheva2018MAGNet}	&$\mathcal{C}_{oo}$	&$\mathcal{U}$	&$\mathcal{N}_{a}$	&$\mathcal{P}_s$	&$\mathcal{E}$	&$\mathcal{V}_u$	&$\mathcal{P}_l$	&$\mathcal{D}$	&$\mathcal{DL}$	
\\\midrule
MADDPG-M \cite{Kilinc2018Noisy}	&$\mathcal{C}_{oo}$	&$\mathcal{U}$	&$\mathcal{A}$	&$\mathcal{I}_c$	&$\mathcal{E}$	&$\mathcal{V}_e$	&$\mathcal{PV}$	&$\mathcal{R}_e$	&$\mathcal{DL}$	\\\midrule
IC3Net \cite{Singh2019IC3Net}	&$\mathcal{C}_{oo}$/$\mathcal{C}_{om}$/$\mathcal{M}$	&$\mathcal{U}$	&$\mathcal{A}$	&$\mathcal{I}_c$	&$\mathcal{E}$	&$\mathcal{V}_e$	&$\mathcal{P}_l$	&$\mathcal{R}_e$	&$\mathcal{CTDE}_{ps}$	\\\midrule
MD-MADDPG \cite{Pesce2020MDMADDPG}	&$\mathcal{C}_{oo}$	&$\mathcal{U}$	&$\mathcal{P}$	&$\mathcal{F}_c$	&$\mathcal{E}$	&$\mathcal{V}_u$	&$\mathcal{P}_l$	&$\mathcal{D}$	&$\mathcal{CTDE}_{c}$	\\\midrule
SchedNet \cite{Kim2019SchedNet}	&$\mathcal{C}_{oo}$	&$\mathcal{L}_b$	&$\mathcal{A}$	&$\mathcal{G}_c$	&$\mathcal{E}$	&$\mathcal{V}_e$	&$\mathcal{P}_l$	&$\mathcal{R}_e$	&$\mathcal{CTDE}_{ip}$	\\\midrule
DCC-MD \cite{Kim2019MessageDropout}	&$\mathcal{C}_{oo}$	&$\mathcal{U}$	&$\mathcal{A}$	&$\mathcal{F}_c$	&$\mathcal{E}$	&$\mathcal{V}_u$	&$\mathcal{V}_l$	&$\mathcal{D}$	&$\mathcal{DL}$	\\\midrule
Agent-Entity Graph \cite{Agarwal2020AEG}	&$\mathcal{C}_{oo}$	&$\mathcal{U}$	&$\mathcal{N}_{a}$	&$\mathcal{P}_s$	&$\mathcal{E}$	&$\mathcal{V}_u$	&$\mathcal{PV}$	&$\mathcal{D}$	&$\mathcal{DL}$	\\\midrule
VBC \cite{Zhang2019VBC}	&$\mathcal{C}_{oo}$	&$\mathcal{L}_b$	&$\mathcal{A}$	&$\mathcal{P}_s$	&$\mathcal{E}$	&$\mathcal{V}_e$	&$\mathcal{V}_l$	&$\mathcal{D}$	&$\mathcal{CTDE}_{ps}$	\\\midrule
NDQ \cite{Wang2020NDQ}	&$\mathcal{C}_{oo}$/$\mathcal{M}$	&$\mathcal{L}_b$	&$\mathcal{A}$	&$\mathcal{P}_s$	&$\mathcal{E}$	&$\mathcal{V}_e$	&$\mathcal{V}_l$	&$\mathcal{R}_g$	&$\mathcal{CTDE}_{ps}$	\\\midrule
IMAC \cite{Wang2020IMAC}	&$\mathcal{C}_{oo}$	&$\mathcal{L}_b$	&$\mathcal{P}$	&$\mathcal{F}_c$	&$\mathcal{E}$	&$\mathcal{V}_u$	&$\mathcal{P}_l$	&$\mathcal{R}_g$	&$\mathcal{CTDE}_{ip}$	\\\midrule
GA-Comm \cite{Liu2020G2ANet}	&$\mathcal{C}_{oo}$	&$\mathcal{U}$	&$\mathcal{P}$	&$\mathcal{G}_c$	&$\mathcal{E}$	&$\mathcal{V}_u$	&$\mathcal{P}_l$	&$\mathcal{D}$	&$\mathcal{CTDE}_{ps}$	\\\midrule
Gated-ACML \cite{Mao2020GatedACML}	&$\mathcal{C}_{oo}$	&$\mathcal{L}_b$	&$\mathcal{P}$	&$\mathcal{I}_c$	&$\mathcal{E}$	&$\mathcal{V}_u$	&$\mathcal{P}_l$	&$\mathcal{S}_p$	&$\mathcal{CTDE}_{ip}$	\\\midrule
Bias \cite{Biases2019Eccles}	&$\mathcal{C}_{oo}$	&$\mathcal{L}_b$	&$\mathcal{A}$	&$\mathcal{F}_c$	&$\mathcal{E}$	&$\mathcal{V}_e$	&$\mathcal{PV}$	&$\mathcal{R}_g$	&$\mathcal{DL}$	\\\midrule
LSC \cite{Sheng2020LSC}	&$\mathcal{C}_{oo}$/$\mathcal{M}$	&$\mathcal{U}$	&$\mathcal{N}_{a}$	&$\mathcal{I}_c$	&$\mathcal{E}$	&$\mathcal{V}_u$	&$\mathcal{V}_l$	&$\mathcal{R}_e$	&$\mathcal{CTDE}_{ps}$	\\\midrule
Diff Discrete\cite{Freed2020UnknownNoise}	&$\mathcal{C}_{oo}$	&$\mathcal{C}_m$	&$\mathcal{A}$	&$\mathcal{F}_c$	&$\mathcal{E}$	&$\mathcal{V}_e$	&$\mathcal{P}_l$	&$\mathcal{D}$	&$\mathcal{CTDE}_{ps}$	\\\midrule
NeurComm \cite{Chu2020NeurComm}	&$\mathcal{C}_{oo}$	&$\mathcal{U}$	&$\mathcal{N}_{a}$	&$\mathcal{P}_s$	&$\mathcal{I}$	&$\mathcal{V}_u$	&$\mathcal{PV}$	&$\mathcal{D}$	&$\mathcal{DL}$	\\\midrule
IP \cite{Qu2020IP}	&$\mathcal{C}_{oo}$	&$\mathcal{U}$	&$\mathcal{N}_{a}$	&$\mathcal{P}_s$	&$\mathcal{I}$	&$\mathcal{V}_u$	&$\mathcal{P}_l$	&$\mathcal{D}$	&$\mathcal{DL}$	\\\midrule
I2C \cite{Ding2020I2C}	&$\mathcal{C}_{oo}$	&$\mathcal{U}$	&$\mathcal{A}$	&$\mathcal{I}_c$	&$\mathcal{E}$	&$\mathcal{V}_u$	&$\mathcal{P}_l$	&$\mathcal{S}_p$	&$\mathcal{CTDE}_{ps}$	\\\midrule
IS \cite{Kim2021IS}	&$\mathcal{C}_{oo}$	&$\mathcal{U}$	&$\mathcal{A}$	&$\mathcal{F}_c$	&$\mathcal{I}$	&$\mathcal{V}_e$	&$\mathcal{P}_l$	&$\mathcal{D}$	&$\mathcal{CTDE}_{c}$	\\\midrule
ETCNet \cite{Hu2020ETCNet}	&$\mathcal{C}_{oo}$	&$\mathcal{L}_b$	&$\mathcal{A}$	&$\mathcal{I}_c$	&$\mathcal{E}$	&$\mathcal{V}_e$	&$\mathcal{P}_l$	&$\mathcal{R}_e$	&$\mathcal{CTDE}_{ps}$	\\\midrule
\newpage
\\\midrule
Variable-length Coding \cite{Freed2020Length}	&$\mathcal{C}_{oo}$	&$\mathcal{L}_b$	&$\mathcal{A}$	&$\mathcal{F}_c$	&$\mathcal{E}$	&$\mathcal{V}_e$	&$\mathcal{P}_l$	&$\mathcal{D}$	&$\mathcal{CTDE}_{ps}$	\\\midrule
TMC \cite{Zhang2020TMC}	&$\mathcal{C}_{oo}$	&$\mathcal{L}_b$	&$\mathcal{A}$	&$\mathcal{P}_s$	&$\mathcal{E}$	&$\mathcal{V}_u$	&$\mathcal{V}_l$	&$\mathcal{D}$	&$\mathcal{CTDE}_{ps}$	\\\midrule
HAMMER \cite{Gupta2021HAMMER}	&$\mathcal{C}_{oo}$	&$\mathcal{U}$	&$\mathcal{P}$	&$\mathcal{F}_c$	&$\mathcal{E}$	&$\mathcal{V}_u$	&$\mathcal{P}_l$	&$\mathcal{R}_e$	&$\mathcal{CTDE}_{ps}$	\\\midrule
MAGIC \cite{Niu2021MAGIC}	&$\mathcal{C}_{oo}$/$\mathcal{M}$	&$\mathcal{U}$	&$\mathcal{P}$	&$\mathcal{G}_c$	&$\mathcal{E}$	&$\mathcal{V}_u$	&$\mathcal{PV}$	&$\mathcal{D}$	&$\mathcal{CTDE}_{ps}$	\\\midrule
FlowComm \cite{Du2021FlowComm}	&$\mathcal{C}_{oo}$	&$\mathcal{U}$	&$\mathcal{N}_{a}$	&$\mathcal{G}_c$	&$\mathcal{E}$	&$\mathcal{V}_e$	&$\mathcal{P}_l$	&$\mathcal{D}$	&$\mathcal{CTDE}_{ps}$	\\\midrule
AE-Comm \cite{Autoencoders2021Lin}	&$\mathcal{C}_{oo}$	&$\mathcal{L}_b$	&$\mathcal{A}$	&$\mathcal{F}_c$	&$\mathcal{E}$	&$\mathcal{V}_e$	&$\mathcal{PV}$	&$\mathcal{R}_g$	&$\mathcal{DL}$	\\\midrule
GAXNet \cite{Yun2021GAXNet}	&$\mathcal{C}_{oo}$	&$\mathcal{U}$	&$\mathcal{N}_{a}$	&$\mathcal{P}_s$	&$\mathcal{E}$	&$\mathcal{V}_u$	&$\mathcal{P}_l$	&$\mathcal{D}$	&$\mathcal{CTDE}_{ip}$	\\\midrule
DCSS \cite{Discrete2021Tucker}	&$\mathcal{C}_{oo}$	&$\mathcal{C}_m$	&$\mathcal{P}$	&$\mathcal{F}_c$	&$\mathcal{E}$	&$\mathcal{V}_e$	&$\mathcal{PV}$	&$\mathcal{D}$+$\mathcal{S}_p$	&$\mathcal{CTDE}_{ip}$	\\\midrule
R-MACRL \cite{Xue2021RMACRL} &$\mathcal{C}_{om}$	&$\mathcal{C}_m$	&$\mathcal{A}$	&$\mathcal{F}_c$	&$\mathcal{E}$	&$\mathcal{V}_u$	&$\mathcal{P}_l$	&$\mathcal{S}_p$	&$\mathcal{DL}$	\\\midrule
MAIC \cite{Yuan2022MAIC}	&$\mathcal{C}_{oo}$	&$\mathcal{L}_b$	&$\mathcal{A}$	&$\mathcal{P}_s$	&$\mathcal{E}$	&$\mathcal{V}_u$	&$\mathcal{V}_l$	&$\mathcal{R}_g$	&$\mathcal{CTDE}_{ps}$	\\\midrule
FCMNet \cite{Wang2022FCMNet}	&$\mathcal{C}_{oo}$	&$\mathcal{U}$	&$\mathcal{A}$	&$\mathcal{F}_c$	&$\mathcal{E}$	&$\mathcal{V}_u$	&$\mathcal{PV}$	&$\mathcal{D}$	&$\mathcal{CTDE}_{ps}$	\\\bottomrule

\end{longtable}
% \end{small}

In addition to these findings, the evaluation metrics used in Comm-MADRL research are of significant interest. It is noteworthy that existing works have been evaluated across various platforms and games, employing different metrics to assess performance. Crucially, Comm-MADRL studies often use varying settings of experiments, such as the number of agents or the use of parameter sharing. These settings can make it challenging to fairly compare the relative strengths and limitations of algorithms based on their performance outcomes \cite{Papoudakis2021Benchmarking}. We have identified four evaluation metrics commonly used in Comm-MADRL studies as follows:

\begin{itemize}[leftmargin=*]
\item \textbf{Reward-based}: This metric employs the converged return or average rewards per episode or time step to demonstrate the profit gained by agents.

\item \textbf{Win or Fail Rate}: This metric calculates the percentage that agents achieve their goal or fail the game during learning. It is often used in episodic tasks.

\item \textbf{Steps Taken}: This metric evaluates the number of time steps learned to reach the goal. It is often used in episodic tasks and essential in scenarios where time efficiency is key.

\item \textbf{Communication Efficiency}: This metric evaluates how much communication resource has been used, such as the frequency of communication between agents.

\item \textbf{Emergence Degree}: Originating from the field of emergent language, this metric evaluates and detects the emergence of language \cite{Bogin2018Consistent,Lowe2019Pitfalls}. It is often used in learning tasks with emergent language. \emph{Positive signaling} and \emph{positive listening} are two common approaches. Positive signaling measures the correlation between a message and the sender’s observation or intended action. Positive listening assesses the impact of an observed message on the receiver's beliefs or behavior.
\end{itemize}

\begin{figure}[t]
 \centering
 \includegraphics[width=0.7\linewidth,height=0.35\linewidth]{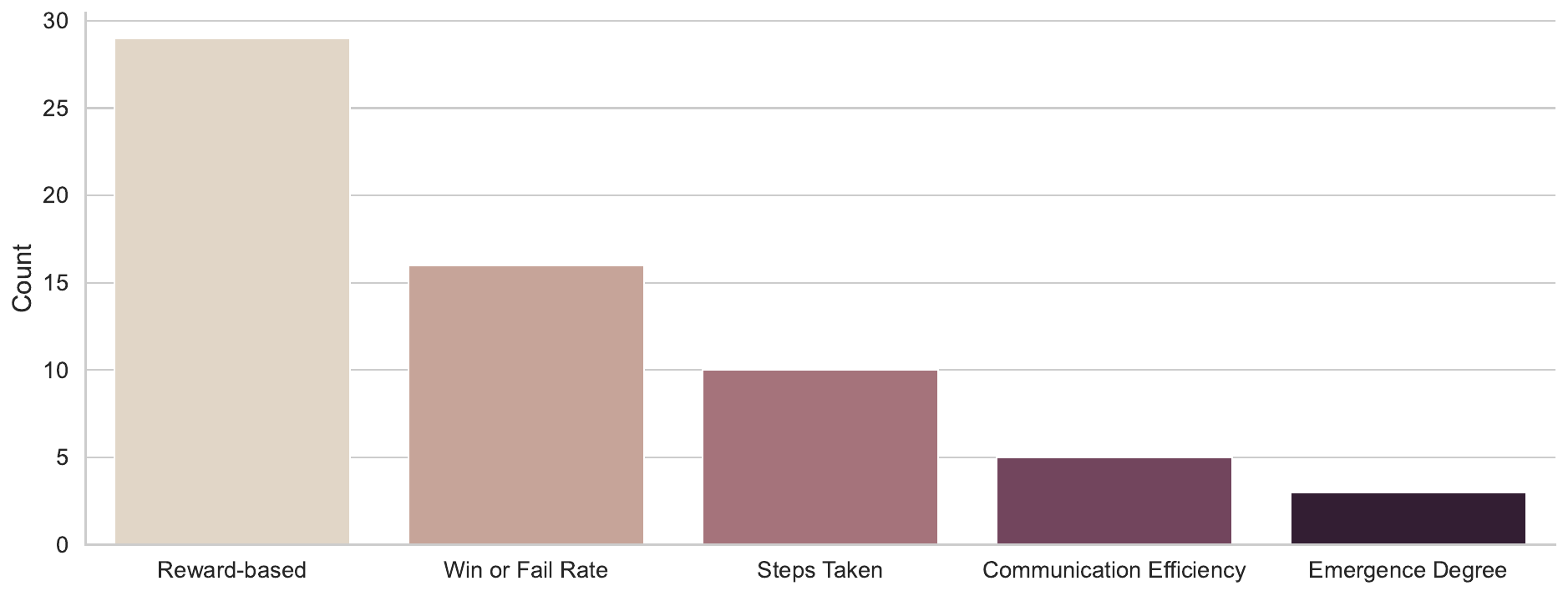}
 \caption{The Statistics of Evaluation Metrics in existing Comm-MADRL systems.}
 \label{fig:statistics}
\end{figure}

We have analyzed the number of times that the above performance metrics are used in existing Comm-MADRL studies, as illustrated in Figure \ref{fig:statistics}. It is shown that the metric of communication efficiency has not been extensively used in the literature, requiring further investigation into the use of communication resources in Comm-MADRL approaches. The Emergence Degree metric, intended to measure whether a language is emergent, is primarily utilized in emergent language studies. Nonetheless, this metric can also yield significant insights for other Comm-MADRL systems. By analyzing the correlation between communication and the observations and behaviors of both senders and receivers, we could obtain a deeper understanding and explanation of communication for Comm-MADRL.

In the next section, inspired by the proposed dimensions, we demonstrate the potential for discovering new ideas through our survey. We identify several possible research directions that jointly explore multiple dimensions, aiming to bridge the gaps in current works.

\subsection{Research Directions}

Comm-MADRL is a young but rapidly enlarging field. There are still lots of possibilities to develop new Comm-MADRL systems. Our proposed dimensions encapsulate several aspects of Comm-MADRL, from which we can identify new research directions. Therefore, we showcase four research directions motivated by leveraging the possible combinations of dimensions and the extensions of corresponding categories. We also point out further challenges for Comm-MADRL.

\subsubsection{Multimodal Communication}

A versatile robot can hear by sound sensors, read text or talk with human partners. Intelligent agents may be surrounded by different data sources and act based on multimodal input. By jointly considering the dimensions of communicatee type and communicated messages, we can imagine a fertile scenario where communication is not limited to images or handcrafted features but encompasses multimodal data, such as speech, videos, and text from humans or domestic robots, to prosper applications like smart home. To the best of our knowledge, existing works in Comm-MADRL do not consider communicating multimodal data or encoding them. Recent works often use encoded images as messages, which only cover visually-based applications. Therefore, we believe exploring multimodal communication represents a promising research direction and introduces several challenges that need to be addressed. In multimodal communication, agents have to coordinate heterogeneous modalities and encode various types of information into messages. A possible solution is to use separate channels to communicate specific modalities, while agents must decide on the right channel to communicate and merge data from different channels. A more efficient way is to learn a joint representation of multimodal observations and communicate on one channel. Due to the progress of Multimodal machine learning \cite{Baltrusaitis2019Multimodal}, we can bring ideas from this area to equip agents with the ability to create a single representation of multimodal data. Nevertheless, it is unclear how the solutions from Multimodal machine learning can be extended to multi-agent reinforcement learning. Poklukar et al. \cite{Poklukar2022Multimodal} propose learning an aligned representation from multiple modalities, although their tests are conducted in a single-agent reinforcement learning task. The multi-agent scenarios, however, may need to consider the individual abilities and preferences of different agents. For example, a voice-activated agent may favor voice data for interaction, while a monitoring agent may only access video data. Therefore, in multi-agent settings, agents need to align their individual preferences regarding multimodality when learning a joint representation of the multimodal data. Another crucial technical issue is how to represent multimodal messages in low-dimensional vectors without losing essential information from each modality, as Comm-MADRL systems often consider reducing communication costs. Eventually, we expect the progress of multimodal communication will benefit human-agent interaction and diverse communicating agents.

The emergence of new research works would introduce new categories under each dimension. For example, with developments in multimodal communication, we can extend the categories of communicated messages with speech, image, text, and video data. Nevertheless, our proposed dimensions can be adaptive and robust to cover new Comm-MADRL research in this direction.

\subsubsection{Structural Communication}

Through the Internet, electronic devices like routers can process and transmit information. On social media, chatbots can be community members \cite{Seering2019Chatbots,Seering2020Gaming}, to engage in conversations with users and share information/opinions. In those large-scale multi-agent systems \cite{Folstad2017Chatbots,Choudhury2004marp}, agents may belong to different groups, where their relationships can be complicated. For example, local area networks create boundaries of communication and interaction between devices. Chatbots may not be able to reach some users because of limited permission or the lack of friendship relations. These restricted connectivities among agents require more efficient usage of communication structure. Therefore, we think that the research direction focusing on structural communication opens up possibilities for enabling communication among a larger number of agents. In the current literature in Comm-MADRL, ATOC \cite{Jiang2018ATOC} and LSC \cite{Sheng2020LSC} have investigated communication with multiple groups, where agents can only communicate with other agents who belong to the same group. In both approaches, different groups may share common member agents, i.e., bridge agents, which are used to enable information to flow from group to group. However, communication through bridge agents is not targeted and each agent unconsciously shares their information with other groups. In terms of the dimension of controlled goals, agents may have individualized goals and require collaboration with a specific set of agents. Therefore, an important future direction of structural communication is to send critical information and opinions to target agents. For example, agent 1 may observe the goal location of agent 2 while they belong to different groups. If agent 3 happens to be a common friend of agent 1 and 2, agent 1 can actively send the goal information to agent 2 with the help of agent 3. If communication is costly and information is private, agents need to make thoughtful decisions about which bridge agents to be used to find the shortest and safe path to reach targeted agents. At the same time, bridge agents need to agree on the communication path to transmit information successfully. If a complex and hierarchical friendship network is identified, another important question is how to prioritize and schedule different communication paths to make communication fluent. Regarding communicated messages, agents need to build a common protocol with targeted agents so that information can be encoded and decoded successfully. As a result, agents can more actively utilize the communication structure among agents to achieve better collaboration and agreements.

\subsubsection{Robust Centralized Unit}

Robustness has been widely considered in the field of reinforcement learning \cite{Pinto2017Adversarial,Pattanaik2018Robust}, where an agent needs to cope with disturbances in learning in order to achieve a robust policy that can generalize under changes in training/test data. In MARL, agents’ policies can be sensitive to environmental noise or malicious intentions of opponents, and thus robust policies are required \cite{Li2019Robust,Zhang2020Robust}. With communication, opponents may produce malicious messages, implying adversary intentions. Preventing malicious messages is important in non-cooperative settings as adversary agents may manipulate communication to achieve their own goals at the expense of other agents' benefits. Existing works on Comm-MADRL, such as $\mathfrak{R}$-MACRL \cite{Xue2021RMACRL}, have investigated how to detect adversary information and reconstruct original messages. However, as we discussed in the dimensions of communicatee type and training paradigm, proxy and critics are often centralized and gather information from all agents. Robustness becomes essential for these centralized units as all agents involved in communication can be misled by polluted feedback, for example, incorrect gradient signals from critics or malicious messages from a proxy. Moreover, malicious messages can easily spread through the (centralized) proxy. Therefore, we think building a robust centralized unit is a promising and underdeveloped direction for safe communication in MADRL, where proxies and critics need to avoid communication being exploited by adversaries or affected by harmful environmental changes. By considering the dimension of communication policy, sender agents can learn a versatile communication policy. For example, the communication policy can be defined to select different encoding protocols for different groups of agents, in case malicious agents may easily find a solution to cheat on a specific encoding protocol. Besides, as malicious or noisy messages can be hidden in the centralized proxy, it is important to figure out which messages are malicious and how to reconstruct the original messages. Nonetheless, developing robust centralized units is vital for reliable and protected Comm-MADRL systems. 

\subsubsection{Learning Tasks with Emergent Language}

In this survey, we have identified the intersection between learning tasks with communication and emergent language in the field of MADRL, which we have called learning tasks with emergent language. We also observed that there is only a limited number of research works concerning this sub-area learning tasks with emergent language, which learns a language while achieving a MADRL task. We believe this area can be further expanded and investigated, by considering several dimensions proposed by our survey. First, the communicated messages, as we discussed earlier, can be encoded into more complex symbolic formats, such as graphs or logical expressions. Existing works in the field only learn how to communicate through atomic symbols or a combination \cite{Populations2018Mordatch,Biases2019Eccles,Discrete2021Tucker}. However, it is important to learn the relation between symbols. For example, symbol A is on the left of symbol B. Those messages can express facts about what agents know or conjecture. Therefore, receivers can quickly adapt their behaviors by successfully decoding the messages. The important question is how to learn both encoding and decoding with complex expressions of messages, which can have a significant number of possibilities. The senders should also properly encapsulate their knowledge and the receivers should reason on the messages correctly. In addition, how complex symbolic formats can emerge in non-cooperative settings is an interesting but unexplored research area. What's more, the combination of complex messages will not be as easy as handling single values or vectors. Therefore, learning together with complicated communication is still challenging.       
 
\subsubsection{Further Challenges}

In the field of Comm-MADRL, there are further challenges. For instance, the design of neural network architectures plays a critical role in performance and communication. A deeper neural network may be effective in some domains while failing in other domains. For example, LSTM is effective in capturing history information while may require much time to train the parameters \cite{Hochreiter1997LSTM,Foerster2017Stabilising}, which could greatly slow down the learning in tasks with high complexity. The choice of architectures and fine-tuning hyperparameters are significant problems of Comm-MADRL. With communication, another crucial issue is the explainability of communicated messages. Emergent language has made a step towards human-like language. However, whether machines communicate in a human-like way and can learn a human-interpretable language is still unclear. A great number of existing works regarding learning tasks with communication seek hidden, deep, and obscure codes for messages \cite{Sukhbaatar2016CommNet,Singh2019IC3Net,Ding2020I2C,Niu2021MAGIC}, which still need to be further interpreted and understood.

\section{Conclusions}

Our survey proposes to classify the literature based on 9 dimensions. These dimensions constitute the basis of designing Comm-MADRL systems. We further categorize existing works under each dimension, where readers can easily compare research works from a unique perspective. Based on those dimensions, we also observe findings through the trend of the literature and identify new research directions by filling the gap among recent works. Our survey concludes that while the number of works in Comm-MADRL is notable and represents significant achievements, communication can be more fruitful and versatile to incorporate non-cooperative settings, heterogeneous players, and many more agents. Agents can communicate information not only from raw image inputs or handcrafted features but also from diverse data sources such as voice and text. Furthermore, we can explore novel metrics to better understand the contribution of communication to the overall learning process. Ultimately, Comm-MADRL can benefit from the MARL community and take advantage of good solutions from MARL.

%%%%%%%%%%%%%%%%%%%%%%%%%%%%%%%%%%%%%%%%%%%%%%%%%%%%%%%%%%%%%%%%%%%%%%%%
%Bibliography
\bibliographystyle{unsrt}  
\bibliography{surveyCommMADRL}

\end{document}